\documentclass[lettersize,journal]{IEEEtran}
\usepackage{amsmath,amsfonts}
\usepackage{algorithmic}
\usepackage{algorithm}
\usepackage{array}

\usepackage{textcomp}
\usepackage{stfloats}
\usepackage{url}
\usepackage{verbatim}
\usepackage{graphicx}
\usepackage{cite}
\usepackage{subfigure}
\usepackage{xcolor}
\usepackage{hyperref} 

\hypersetup{
    colorlinks=true,   
    linkcolor=blue,    
    citecolor=blue,    
    urlcolor=blue      
}
\usepackage{booktabs}

\begin{document}

\title{Resonant Beam Multi-Target DOA Estimation}

\author{Yixuan Guo, Qingwei Jiang, Mingliang Xiong,~\IEEEmembership{Member,~IEEE,} Wen Fang, Mingqing Liu, 

Qingqing Zhang,~\IEEEmembership{Member,~IEEE,} Qingwen Liu,~\IEEEmembership{Senior Member,~IEEE,} and Gang Yan,~\IEEEmembership{Member,~IEEE}
\thanks{Y. Guo is with the Shanghai Research Institute for Intelligent Autonomous Systems, Tongji University, Shanghai 201210, China 
(e-mail: guoyixuan@tongji.edu.cn).

M. Xiong is with the Department of Computer Science and Technology, Tongji University, Shanghai 201804, China (email: mlx@tongji.edu.cn).

Q. Jiang, W. Fang, and Q. Liu is with the College of Electronics and Information Engineering, Tongji University, Shanghai 201804, China
(e-mail: \{jiangqw, wen.fang, qliu\}@tongji.edu.cn).

M. Liu is with the LiFi Research and Development Center, Department of Engineering, University of Cambridge, Cambridge, UK (email: ml2176@cam.ac.uk).

Q. Zhang is with the School of Information Engineering, Zhejiang University of Technology, Hangzhou, Zhejiang 310014, and also with Huzhou Key Laboratory of Information-Energy Convergence and Interconnection, Yangtze Delta Region Institute (Huzhou), University of Electronic Science and Technology of China, Huzhou, Zhejiang 313001, China (email: qingqingzhang@zjut.edu.cn).

G. Yan is with the School of Physics Science and Engineering, Tongji University, Shanghai 200092, China (e-mail: gyan@tongji.edu.cn).
	}
}

\maketitle

\begin{abstract}
With the increasing demand for internet of things (IoT) applications, especially for location-based services, how to locate passive mobile targets (MTs) with minimal beam control has become a challenge. Resonant beam systems are considered promising IoT technologies with advantages such as beam self-alignment and energy concentration. To establish a resonant system in the radio frequency (RF) band and achieve multi-target localization, this paper designs a multi-target resonant system architecture, allowing a single base station (BS) to independently connect with multiple MTs. By employing a retro-directive array, a multi-channel cyclic model is established to realize one-to-many electromagnetic wave propagation and MT direction-of-arrival (DOA) estimation through echo resonance. 
Simulation results show that the proposed system supports resonant establishment between the BS and multiple MTs. This helps the BS to still have high DOA estimation accuracy in the face of multiple passive MTs, and can ensure that the DOA error is less than 1$^\circ$ within a range of 6 meters at a 50$^\circ$ field of view, with higher accuracy than active beamforming localization systems.

\end{abstract}


\section{Introduction}
\begin{figure}[!t]
\centering
\includegraphics[width=0.8\linewidth]{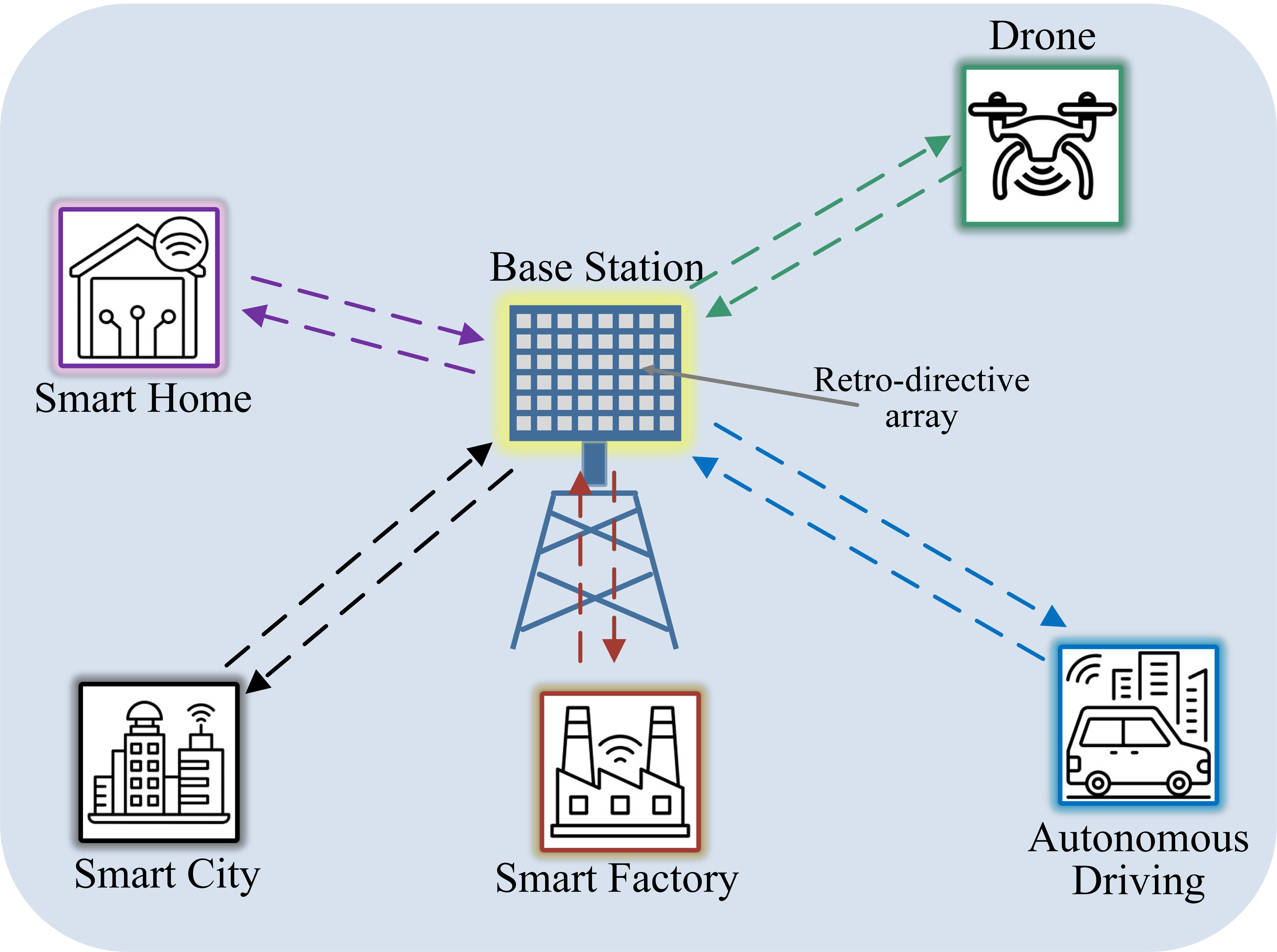}
\caption{Location-based service application scenarios.}
\label{applications}
\end{figure}

\IEEEPARstart{T}{he} rapid development of wireless communication networks has provided a significant boost and broad application prospects for the internet of things (IoT). By revealing, extracting, and utilizing key array propagation parameters, such as the direction of arrival (DOA), IoT devices can more efficiently perceive the location information of themselves and their communication targets \cite{li2017direction}. This not only ensures data transmission capabilities between IoT devices but also significantly reduces power consumption, enabling various smart devices to maintain longer online operation. 
Meanwhile, more accurate location information makes location-based services such as autonomous driving, smart homes, smart cities, and industrial automation more feasible and widespread, as shown in Fig.~\ref{applications}.
Current research on DOA estimation mainly focuses on improving estimation accuracy, reducing complexity, and increasing degrees of freedom through array configuration and algorithm design\cite{chu2023exploration}\cite{7400949}.

In terms of array configuration, researchers design antennas with different shapes and structures to enhance the spatial resolution and signal reception capabilities of the array. For example, \cite{xia2007decoupled} investigates the use of two parallel uniform linear arrays (ULA) for decoupled estimation of two-dimensional angle of arrival (AOA), simplifying the computational complexity. In \cite{10148803}, a coprime planar array (CPPA) is utilized, and methods like matrix completion and sparse matrix recovery are applied to improve the accuracy of DOA estimation. In \cite{sanchez2021gridless}, a gridless multi-dimensional AOA estimation method is developed to improve the estimation accuracy for arbitrary three-dimensional antenna arrays.

In terms of estimation algorithm design, classical high-resolution algorithms such as conventional beamforming techniques (CBF) \cite{bf}, Capon \cite{capon}, multiple signal classification (MUSIC) \cite{gupta2015music} and estimation of signal parameters via rotational invariance techniques (ESPRIT) \cite{ESPRIT} have been extensively studied. These methods utilize the characteristics of signals received by the array to improve the accuracy and resolution of DOA estimation through eigenvalue decomposition or sparse signal reconstruction techniques \cite{oumar2012comparison}. They are suitable for low-power devices and have strong interpretability.  However, in complex environments, such as non-line-of-sight (NLOS) conditions, multipath propagation, and interference, the system performance can be significantly affected. Machine learning-based DOA estimation algorithms can improve adaptability to low signal-to-noise ratios (SNR) conditions and learn complex channel characteristics through large-scale training data, minimizing the impact of environmental interference as much as possible. However, they require high computational resources and have poor interpretability \cite{orlando2023comparative}\cite{akter2021rfdoa}.

However, the aforementioned methods still face challenges when dealing with passive mobile targets (MTs), such as objects or interference sources that lack active signal emission. The passive MTs do not emit signals on their own, relying instead on reflected or scattered signals from the environment for localization. This presents difficulties for traditional arrays and estimation algorithms, as they encounter low SNR and severe multipath effects. Additionally, the position of passive MTs may dynamically change over time, requiring real-time adjustments in estimation strategies, further increasing system complexity.

To address these issues, researchers are exploring DOA estimation methods based on multi-sensor fusion \cite{ma2019multi}, cooperative localization \cite{ma2019multitag}, and reconfigurable intelligent surface (RIS) technologies \cite{keykhosravi2021semi} to enhance the localization accuracy of passive MTs and the overall system performance. However, these methods typically require substantial hardware support or complex beamforming control on the one hand, and on the other hand, the limited power reflected back by passive MTs restricts the improvement of localization accuracy.

Emerging resonant beam systems (RBS) provide inspiration, originally applied in optical wireless energy transfer. In such systems, lasers emitted from the base station (BS), using cat's-eye devices on both the BS and the mobile MT, enable passive long-range energy transfer. The core principle is that during the repeated transmission of light between the two cat's-eye structures, beams with mismatched phase and frequency are canceled, while beams with consistent phase and frequency are reinforced \cite{xiong2020resonant}\cite{zhang2024resonant}. Through multiple iterations, resonance between the BS and the MT can be achieved, enabling highly efficient energy transfer \cite{zhang2018distributed}\cite{liu2016charging}. However, optical systems face limitations due to low photoelectric conversion efficiency, high deployment costs, and challenges in NLOS environments, restricting further development, the \cite{guo2024resonant} and \cite{xia2024millimeter} proposes a resonance system suitable for the millimeter wave band, but this structure has limitations in multi-target scenarios. To locate multiple MTs, it is necessary to achieve multi-channel resonance while tracking and estimating the feature vectors associated with each MT. Therefore, we have redesigned the resonance system in the radio frequency (RF) band, supporting the BS to simultaneously establish resonance beams with multiple MTs based on the different signal strengths of the signals returned by each MT. By analyzing and processing the mixed signals received by the BS array, we have obtained the respective angle information of passive MTs.

The main contributions of this paper are as follows:
\begin{itemize}
\item Considering that existing passive localization techniques often require extensive hardware support or complex beam control and are limited by the weak signal strength of MT reflections, we propose a multi-target resonance system architecture in the RF frequency band. Both the MTs and the BS are equipped with retro-directive arrays (RDA), capable of dynamically adjusting phase based on the received signals. Through multiple iterations, echo resonance is achieved, enabling adaptive beam alignment between the BS and each MT without the need for complex signal processing.

\item To verify the feasibility of the RF resonance-based multi-target passive localization scheme, we developed an analytical model for a multi-target passive localization system. The electromagnetic wave propagation model between each MT and the BS was established based on echo characteristics. Utilizing the MUSIC algorithm, the fused signals from multi-target were processed to extract each MT's angular information, confirming the correctness of the model.

\item Simulation results demonstrate that the proposed system can maintain a high signal reception strength at the BS while performing adaptive beam alignment with multiple passive targets. Consequently, it can achieve an error of less than 1 degree within a range of 6 meters and a 50$^\circ$ field of view, outperforming active localization systems and providing a novel approach for exploring multi-target resonance systems.

\end{itemize}
\section{System Design}

\subsection{Retro-Directive Array}

\begin{figure}
  \centering
        \subfigure[]{
	\includegraphics[width=0.4\linewidth]{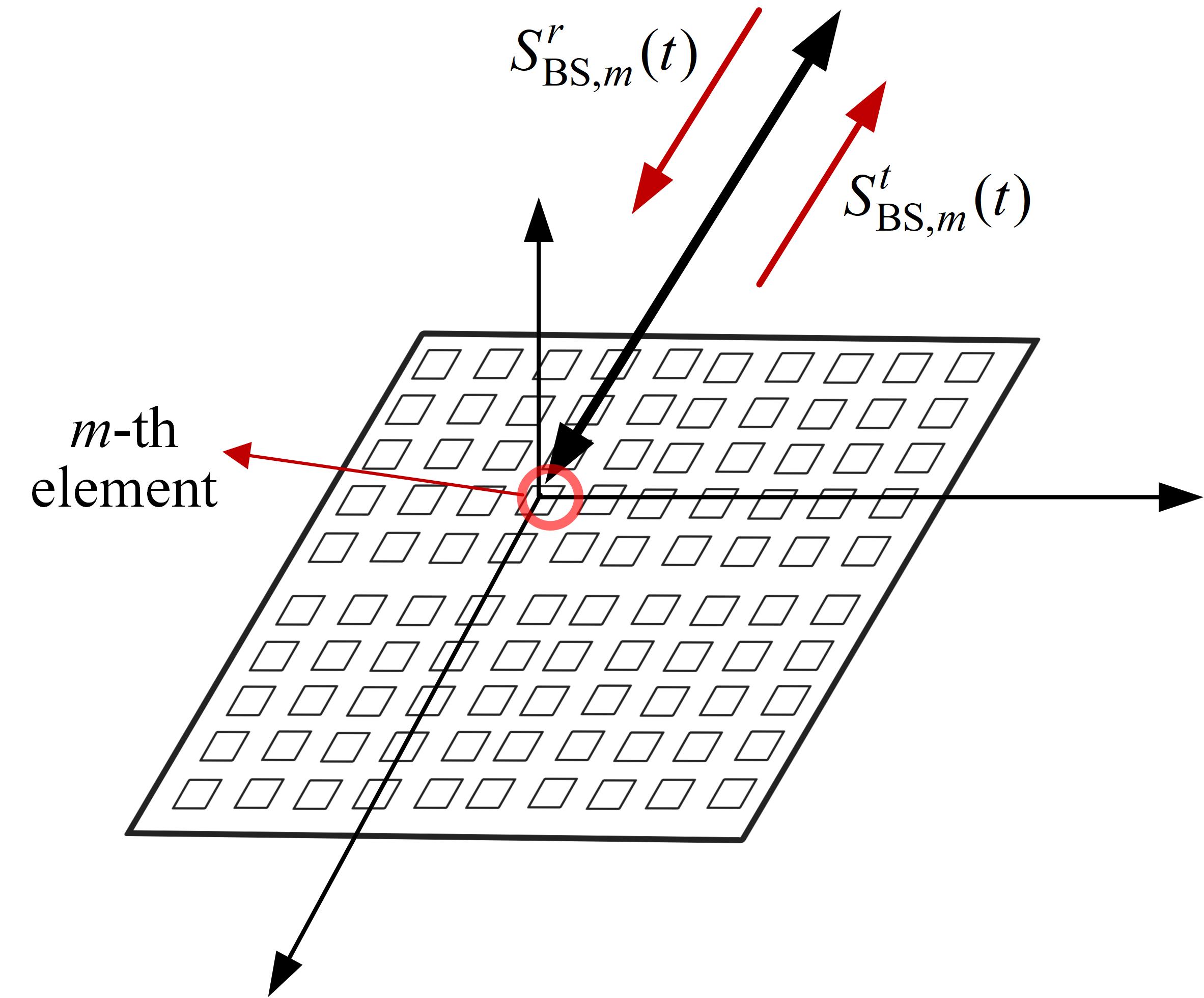}}
 \subfigure[]{
	\includegraphics[width=0.4\linewidth]{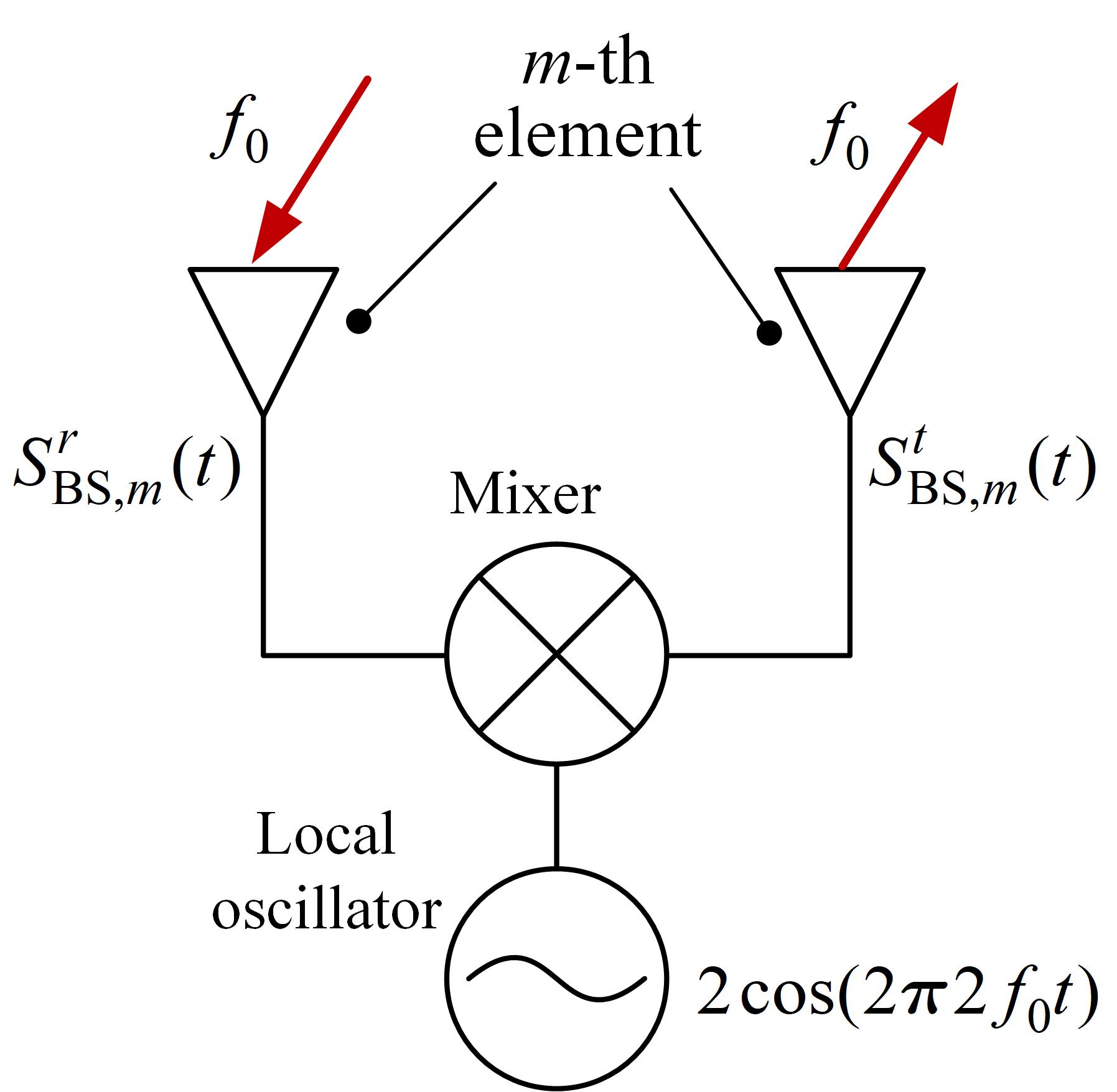}}
\caption{Schematic diagram and principle of retro-directive array.}
\label{RDA}
\end{figure}

As shown in Fig.~\ref{RDA}(a), retro-directive array (RDA) is an antenna system that can automatically transmit received signals back to the signal source in reverse, without the need for prior knowledge of the angle of the incident signal or complex signal processing algorithms \cite{miyamoto2002retrodirective}. Its working mechanism includes two main methods:

i) The Van Atta array consists of paired antennas connected by equal length connecting wires. The received signal will be transmitted through these connection lines and transmitted back to the signal source through the antenna in reverse. Its advantage is a wide bandwidth, but it requires that the received wave must be a plane wave \cite{sharp1960van}\cite{ang2018passive}.

ii) The phase conjugate mixer array achieves reverse transmission through phase conjugate technology, as shown in Fig.~\ref{RDA}(b). After receiving a signal $\cos(2\pi f_0t+\varphi_r)$ with a frequency of $f_0$, the antenna mixes it with a local oscillation signal $2\cos(2\pi2f_0t)$ at time $t$ in the mixer to generate a new signal
\begin{align}
s^\dagger&=\cos(2\pi f_0t+\varphi_r)\cos(2\pi 2f_0t)\\
   &=\frac{1}{2}[\cos(-2\pi f_0t+\varphi_r)+\cos(6\pi f_0t+\varphi_r)].
\end{align}

By filtering out high-frequency components, we obtain the phase conjugate signal $s^*=\cos(-2\pi f_0t+\varphi_r)$, which is then re-transmitted back to the direction of the signal source \cite{allen2003negative}\cite{pon1964retrodirective}. Compared to Van Atta arrays, arrays using phase conjugation do not require strict antenna layouts and can be installed on curved surfaces, making them suitable for more complex application scenarios. Therefore, the phase conjugate mixer array is adopted in this paper.

\subsection{System Workflow}

\begin{figure*}
  \centering
	\includegraphics[width=0.8\linewidth]{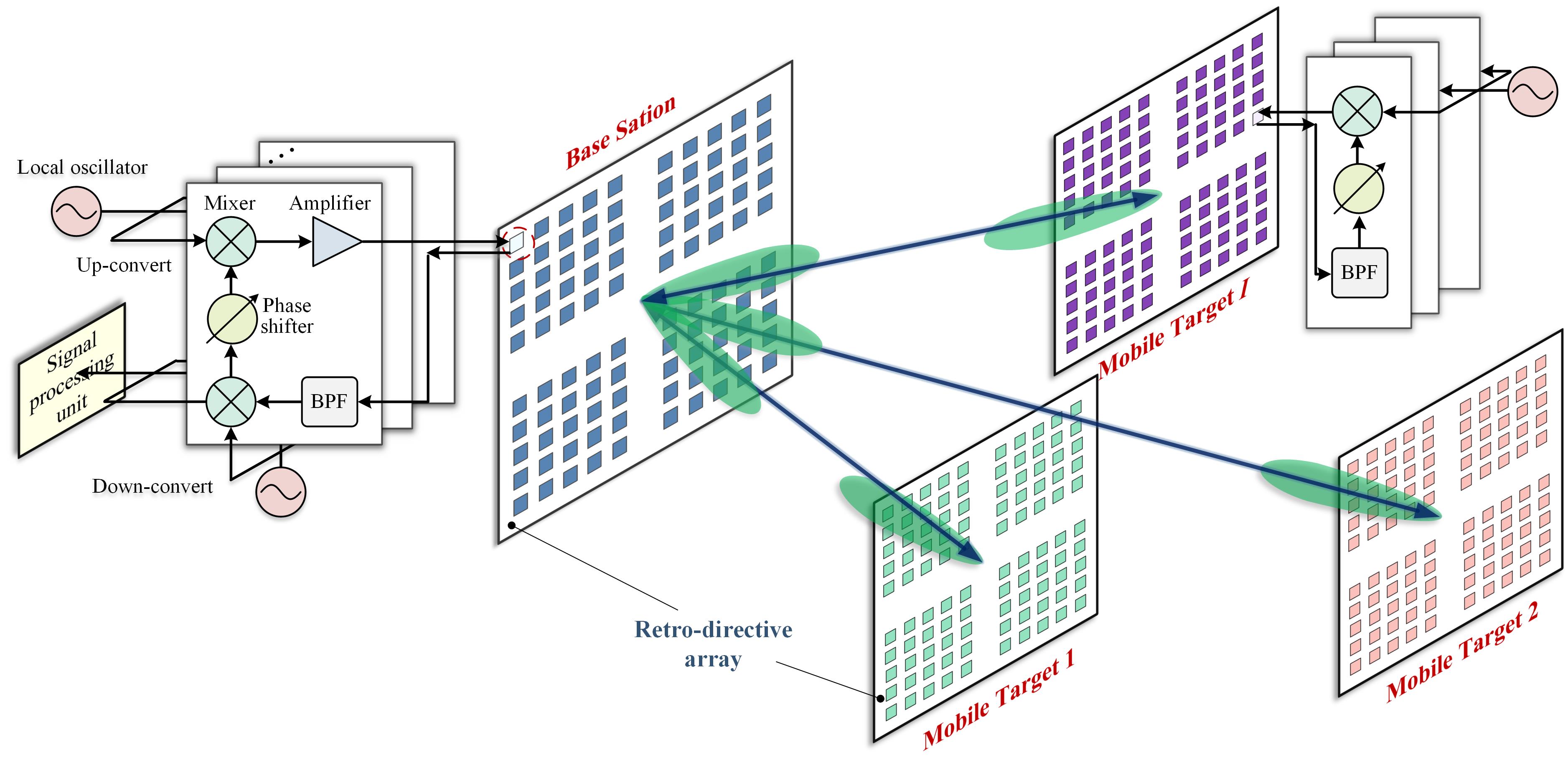}
 \caption{Detailed structure of RF multi-target resonance localization system.}
\label{system_design}
\end{figure*}
As shown in Fig.~\ref{system_design}, the multi-target resonance localization system (MRLS) consists of a BS and $I$ MTs that do not need to actively transmit signals. Both are equipped with RDAs composed of retro-directive antennas. The signal is initially broadcast outward by the BS and, upon reaching the RDA of each MT, it first passes through a band-pass filter (BPF) to remove unwanted frequency components. Then, a phase shifter is used to compensate for any phase errors that may occur during operation. It is important to note that since the MTs are passive, passive phase shifters can be employed \cite{poon2012supporting}. . Subsequently, the signal is fed into a mixer for the aforementioned phase conjugation, generating a return signal that has the same frequency as the received signal and carries the conjugated phase, allowing it to travel back to the BS along the original path \cite{guo2024retrodirective}.

Upon receiving the return signals from the MTs, the BS's RDA also filters the signals through a BPF and then down-converts them using the first mixer and a local oscillator for processing by the signal processing unit. The signal processing unit is responsible for all signal processing tasks and the DOA estimation operations. Following this, the signal undergoes phase shifting and phase conjugation operations to generate the return signal. Unlike the MTs, the local oscillator used for conjugation here needs to have the capability of up-conversion as well. Finally, the signal is amplified by a power amplifier before being transmitted. The power amplifier is responsible for compensating for all possible losses during electromagnetic wave propagation, ensuring that each retransmitted signal is of high power.


\begin{figure}
  \centering
	\includegraphics[width=\linewidth]{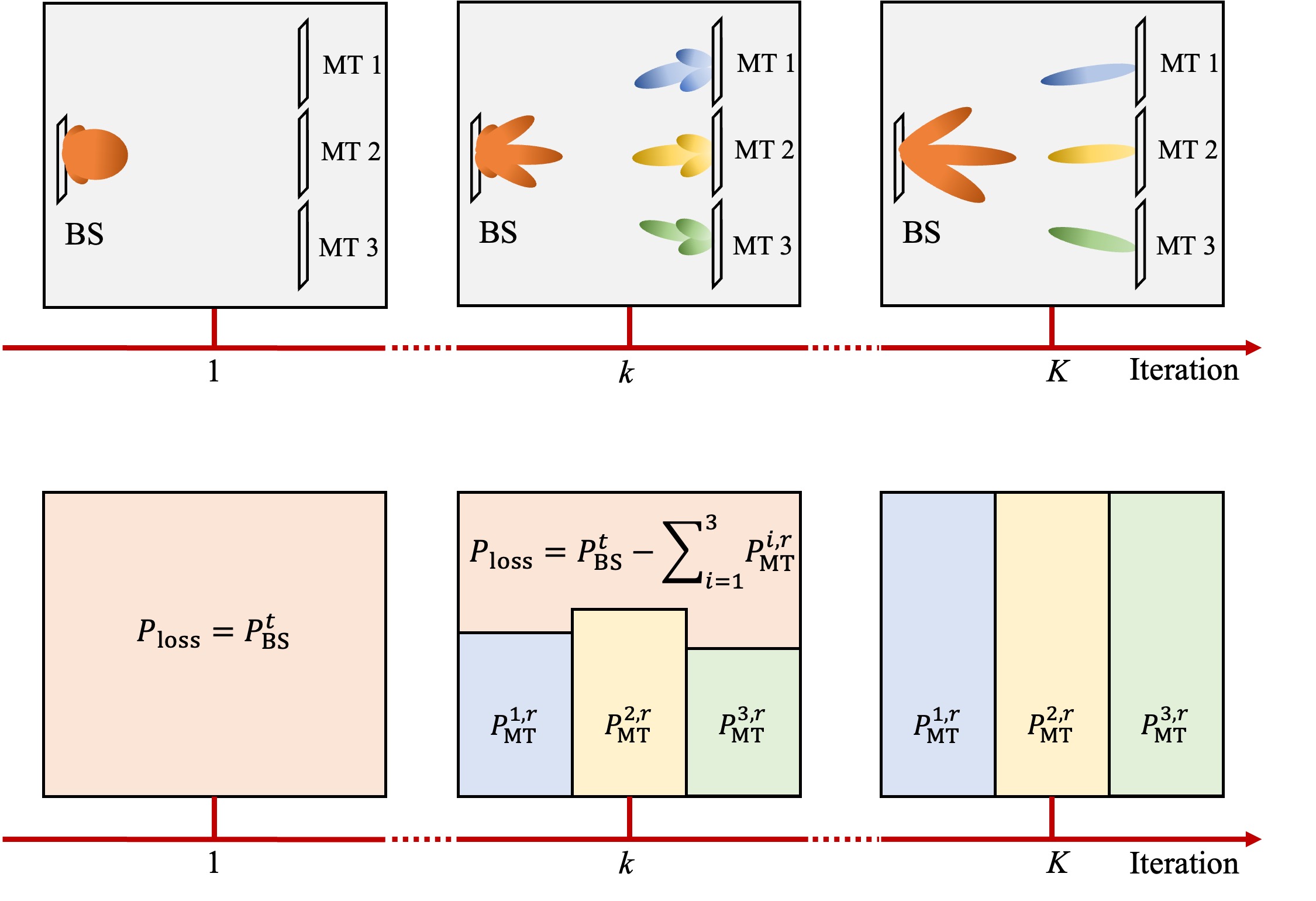}
\caption{The process of resonance formed by echoes between a BS and three MTs. (a) The beam directivity becomes more pronounced with increasing iterations. (b) The system power loss tends to zero as the iteration increases. }
\label{iterations}
\end{figure}

\subsection{Resonance formation process}

Figure \ref{iterations} illustrates the resonance of electromagnetic waves between the BS and the MTs in MRLS as the iteration progresses. Fig.\ref{iterations}(a) shows the variation of the beam during this process, while Fig.\ref{iterations}(b) depicts the variation of system loss during the iteration process.

At the beginning of the first iteration, the BS radiates electromagnetic waves in all directions, and at this stage, the system loss $P_\text{loss}$ is equal to the output power of the BS $P_\text{BS}^t$. When the passive MTs, also equipped with an RDA, receives electromagnetic waves from the BS, it returns waves of the same frequency back to the BS through the RDA. Upon receiving these reflected signals, the BS further transmits electromagnetic waves in specific directions, significantly reducing system losses, which is equal to the BS's transmitted signal power minus the signal power received by all MTs, i.e. $P_\text{loss}=P_\text{BS}^t-\sum^I_{i=1}P_{\text{MT}}^{i,r}$. However, since the phases of the bidirectional electromagnetic waves in space are not yet fully aligned, the electromagnetic waves radiated by both the BS and the target still exhibit strong sidelobes, resulting in relatively low efficiency.

As the iterative process continues, electromagnetic waves with the same phase constructively interfere, while those with different phases destructively interfere. This process continues until the bidirectional electromagnetic waves achieve complete phase alignment, forming a highly adaptive beamforming pattern in space. At this point, the output power of the BS experiences almost no loss, indicating that the system has reached a steady state—where the bidirectional electromagnetic waves establish resonance in space, similar to the formation of a standing wave by bidirectional light beams in a laser cavity.

This phenomenon is fundamentally similar to the analog beamforming mechanism supported by IEEE 802.11ad, which also operates between the transmitter and the receiver \cite{6392842}. However, unlike traditional beamforming, the beamforming in MRLS is self-generated, highly directional, and does not require channel estimation or beam control. Moreover, the user does not need to consume power, making it fundamentally distinct from conventional beamforming approaches.

\section{Mathematical Analysis}
\subsection{Channel Model}
We consider a narrowband multi-user resonance localization system, which consists of a BS equipped with $M$ RDA elements and $I$ MTs equipped with $N$ RDA elements, and satisfies $M \gg I$, that is, the number of BS elements is greater than the MT number. These elements form a uniform planar array (UPA) and operate at the same center frequency. The coordinates of the elements of BS and the $i$-th MT can be expressed as

\begin{equation}
    C_\text{BS} = 
\begin{bmatrix}
(p_{x} - 1) d \\
(p_{y} - 1) d \\
0
\end{bmatrix},
\label{eq1}
\end{equation}

\begin{equation}
C_\text{MT}^i = 
\begin{bmatrix}
(q_{x}^{i} - 1) d \\
(q_{y}^{i} - 1) d \\
Z_{i}
\end{bmatrix},
\end{equation}
where $p_{x}\in\{1,2,\cdots,\sqrt{M}\}$ and $p_{y}\in\{1,2,\cdots,\sqrt{M}\}$ represent the index of the BS RDA element in the $x$-axis and $y$-axis directions, and $q_{x}^{i}\in\{1,2,\cdots,\sqrt{N}\}$ and $q_{y}^{i}\in\{1,2,\cdots,\sqrt{N}\}$ represent the index of the $i$-th MT RDA element in the $x$-axis and $y$-axis directions, $d$ is the distance between adjacent array elements. Initially, the BS broadcasts pulse modulated radiation signals, which are reflected by the MT and returned to the BS after reaching the MT. When the initial phase and amplitude of the signal from the MT are known, the complex expression of the radiation signal from the $i$-th MT can be written as \cite{doan2020doa}

\begin{equation}
S_\text{MT}^{i,t}(t) = \sqrt{2 \mu_0 P_\text{MT}^{i,t}(t)} \cos(\omega t + \varphi_0),
\end{equation}
where $\mu_0$ is the wave impedance \cite{yu2011light}, the $P_\text{MT}^{i,t}(t)$ is the total emitted power of all elements of the $i$-th MT at time $t$, and $\omega = 2 \pi f$ represents the angular frequency of the radiation signal with the carrier frequency $f$. The $\varphi_0$ is the initial carrier phase. Assuming the signal received by the first array element of the BS is known, the received signal matrix at the BS array can be written as \cite{zhang2013high}

\begin{equation}
\mathbf{X}_\text{BS}^r(t) = \mathbf{A}(\theta, \phi)\mathbf{S}_\text{BS}^{r}(t) + \mathbf{P}_\text{N}(t),
\label{eq4}
\end{equation}
where the $\mathbf{S}_\text{BS}^{r}(t) \in \mathbb{C}^{I\times1}$ is the signal received matrix, and
$\mathbf{S}_\text{BS}^{r}=[S_\text{BS}^{1,r}(t),S_\text{BS}^{2,r}(t),...,S_\text{BS}^{I,r}(t)]^T$ is composed of the complex signal received by the BS from the each passive MT at the time $t$,

\begin{equation}
S_\text{BS}^{i,r}(t) = \sqrt{2\mu_0P_\text{BS}^{i,r}(t)} \cos(k l_{mn} - \omega t + \varphi_0 + \varphi{'}),
\label{eq5}
\end{equation}
where the $P_\text{BS}^{i,r}(t)$ is the power of electromagnetic waves radiated from the $i$-th MT to BS, $k=2\pi/\lambda$ is the wavenumber that varies depending on the frequency, the $l_{mn}$ is the distance between the $m$-th BS element and the $n$-th element of the $i$-th MT, $\varphi_0$ is the initial phase, $\varphi'$ is the phase noise.

In eqation (\ref{eq4}), the $\mathbf{P}_\text{N}(t)\in \mathbb{C}^{M\times1}$ represents the additive white Gaussian noise (AWGN) received by the BS array at time $t$, which may be related to noise generated by the MTs  being located or detected internally by the equipment. $\mathbf{A}(\theta, \phi)\in \mathbb{C}^{M\times I}$ is the steering vector matrix \cite{guo2024resonant}, and can be expressed as
\begin{equation}
\mathbf{A}(\theta, \phi) = 
\begin{bmatrix}
    \alpha_{1}^1(\theta_1 \phi_1) & \alpha_{1}^2(\theta_2 \phi_2) & \cdots & \alpha_{1}^I(\theta_I \phi_I) \\
    \alpha_{2}^1(\theta_1 \phi_1) & \alpha_{2}^2(\theta_2 \phi_2) & \cdots & \alpha_{2}^I(\theta_I \phi_I) \\\vdots & \vdots & \ddots & \vdots \\
    \alpha_{M}^1(\theta_1 \phi_1) & \alpha_{M}^2(\theta_2 \phi_2) & \cdots & \alpha_{M}^I(\theta_I \phi_I)
\end{bmatrix}
\end{equation}
where $\alpha_{m}^i$ is determined by the DOA of the signal and the position of the BS elements. Specifically, it is determined by the DOA of the signal reaching the $m$-th array element of the BS and the relative position of that array element to the coordinate origin \cite{ma2021computation}. Combining equation (\ref{eq1}), the $\alpha_{m}^i(\theta_i, \phi_i)$ can be expressed as
\begin{equation}
\alpha_{m}^i(\theta_i, \phi_i) = e^{k\Delta\varphi},\nonumber
\end{equation}
\begin{align}
\Delta\varphi =  &\lfloor\frac{m}{\sqrt{M}}-1\rfloor d\cos\theta_i\cos\phi_i +\\
& \{(m \mod \sqrt{M})-1\}d\cos\theta_i\sin\phi_i.
\end{align}

Note that the phase difference $\Delta\varphi$ is different from that in (\ref{eq5}), as the phase difference here is caused by the delay in the signals reaching different elements within the array. In formula (\ref{eq5}), the phase difference is caused by the signal reflections at different array elements, but they are received at the same time. 
The phase noise $\varphi{'}$ in (\ref{eq5}) follows a distribution of $\varphi{'}\sim\mathcal{N}(0, \sigma_{\varphi'}^2)$, where $\sigma_{\varphi'}^2$ is the variance of the phase noise. The electromagnetic waves experience phase noise during transmission between BS and MTs, especially in internal circuits of the system. Vibrations, thermal noise, scattering noise, and non-linear effects of components inside the circuit may also cause phase noise. These noise sources are amplified and fed back, resulting in phase variations in both the received and transmitted signals.

In numerical values, the phase noise can be studied by subjecting each processing unit to independently and identically distributed (i.i.d.) Gaussian noise with zero mean. The typical representation of phase noise is power spectral density, which indicates the power density at a certain offset frequency. To obtain the phase jitter in the time domain, we need to convert the power spectral density of phase noise $W_{\varphi'}(f)$ into linear scale $L(f) = 10^{\frac{W_{\varphi'}(f)}{10}}$, and the variance of phase noise can be expressed as
\begin{equation}
    \sigma_{\varphi'}^2 = 2 \int_{f_{\text{min}}}^{f_{\text{max}}} L(f) \, df,
\end{equation}
where the $f_{\text{max}}$ and $f_{\text{min}}$ are the upper and lower bounds of the offset frequency.

The SNR of the system, representing the ratio of the signal power to the total noise power received at the BS, is given by
\begin{equation}
\text{SNR} = \frac{\sum_{i=1}^{I}P_\text{BS}^{i,r}(t)}{\sum \mathbf{P}_\text{N}(t)}
\end{equation}

\subsection{Power Cycle Model}
Since MRLS is a passive system, the electromagnetic waves received by BS are actually initially emitted by BS, then received and reflected back to BS by MT. Therefore, to calculate the power received by BS, it is first necessary to determine the power density radiated from the $m$-th BS antenna to the $n$-th antenna of the $i$-th MT \cite{wu2022accurate}

\begin{equation}
W^i_{\text{BS},m \rightarrow\text{MT},n}(t) = \frac{P_{\text{BS}}^{i,t}(t) G_{\text{BS},m} G_{\text{MT},n}}{4 \pi l_{mn}^2},
\end{equation}
\begin{figure*}[!t]
    \centering
    \begin{equation}
E^i_{\text{BS},m-\text{MT},n}(t) = \sqrt{2 \mu_0 W^i_{\text{BS},m \rightarrow\text{MT},n}(t)} 
\cos \left( kl_{mn} - \omega t + \varphi_m + \varphi{'} \right)
\label{eq11}
    \end{equation}
    \begin{align}
\nonumber E^i_{\text{BS}\rightarrow\text{MT},n}(t) &= \sum_{m=1}^{M} \sqrt{2\mu_0 W^i_{\text{BS},m-\text{MT},n}(t)} \cos\left(kl_{mn} - \omega t + \varphi_m + \varphi{'}\right)\\
&=\sum_{m=1}^{M} \sqrt{\frac{P_\text{BS}^{i,t}(t) G_{\text{BS},m} G_{\text{MT},n} \mu_0}{2 \pi l_{mn}^2}} \cos \left( k l_{mn} - \omega t + \varphi_m + \varphi{'} \right)
\label{eq12}
\end{align}
\begin{equation}
E^i_{\text{BS} \rightarrow\text{MT}} = \sum_{m=1}^{M} \sum_{n=1}^{N} \frac{P_{\text{BS}}^{i,t}{(t)} G_{\text{BS},m} G_{\text{MT},n} \mu_0}{2 \pi l_{mn}^2} e^{j \left( kl_{mn} + \varphi_m + \varphi{'} \right)}
\label{eq13}
\end{equation}
\begin{align}
   P_{\text{MT}}^{i,r} = W^i_{\text{BS}\rightarrow\text{MT}} A^i_{\text{eff}} = \frac{\lambda^2}{8 \pi \mu_0} \left( E^i_{\text{BS} \rightarrow\text{MT}}\right)^2= \frac{\lambda^2}{16 \pi^2} \left( \sum_{m=1}^{M} \sum_{n=1}^{N} \frac{P_{\text{BS}}^{i,t} G_{\text{BS},m} G_{\text{MT},n}}{l_{mn}} e^{j(kl_{mn} + \varphi_m + \varphi{'})} \right)^2
   \label{eq14}
\end{align}
        \rule{\linewidth}{0.5pt} 
\end{figure*}

\noindent 
 where $P_{\text{BS}}^{i,t}(t)$ and $G_{\text{BS},m}$ represent the output power and antenna gain of the $m$-th BS antenna, and $G_{\text{MT},n}$ is the gain of the $n$-th MT antenna. The antenna gain mainly depends on the elevation angles $\theta$, reaching its maximum when $\theta=0$, i.e., $G = G_{\text{max}} \cos(\theta)$. The $l_{mn}$ is the distance between the $m$-th BS antenna and the $n$-th MT antenna. Based on the relationship between the electric field intensity and the power density in free space propagation, the electric field radiated from the $m$-th BS antenna to the $n$-th MT antenna can be expressed as (\ref{eq11}), and the total electric field received at the $n$-th MT antenna from the entire BS array depicted in (\ref{eq12}).

Since the MRLS can be regarded as a linear time invariant system \cite{xia2024millimeter}\cite{xiong2021retro} and the time $t$ does not affect the calculation of the field intensity and power superposition of electromagnetic waves in MRLS, it can be removed. By superimposing the field intensity of each antenna and representing the phase in complex form, we can obtain the complex expression of the electric field intensity radiated from the BS array to the MT array as (\ref{eq13}).

The power received at the $n$-th antenna of the $i$-th MT is determined by the effective receiving area $A^i_{\text{eff}}$ and the power density on its surface. Therefore, combining equation (\ref{eq13}), the total power received at the MT array from the BS array can be expressed as (\ref{eq14}), where the phase that reaches the $n$-th antenna of MT is
\begin{equation}    \varphi_{\text{MT},n}^{i,r}=kl_{mn}+\varphi_m+\varphi{'}.
\end{equation}

After receiving electromagnetic waves from BS, MT returns the received waves to BS in a certain proportion $\beta$ through the built-in power divider and phase conjugation circuit, i.e., $P_\text{MT}^{i,t}=\beta \cdot P_\text{MT}^{i,r}$, and after the conjugate operation by the RDA array, the phase of the radiated electromagnetic waves by MT can be expressed as

\begin{equation}
\varphi_{\text{MT},n}^{i,t} =(\varphi_{\text{MT},n}^{i,r})^*= -kl_{mn} - \varphi_m + \varphi{'}.
\end{equation}

\begin{figure*}[!t]
    \centering
\begin{equation}
P_{\text{MT}}^{i,t} = \frac{\beta \lambda^2}{16 \pi^2} \left( \sum_{m=1}^{M} \sum_{n=1}^{N} \frac{P_{\text{BS}}^{i,t} G_{\text{BS},m} G_{\text{MT},n}}{l_{mn}} e^{-j(k l_{mn} + \varphi_m)+\varphi{'}} \right)^2
\label{eq17}
\end{equation}
\begin{equation}
P_{\text{BS}}^{i,r} = \frac{\lambda^2}{16 \pi^2} \left( \sum_{m=1}^{M} \sum_{n=1}^{N} \frac{P_{\text{MT}}^t G_{\text{BS},m} G_{\text{MT},n}}{l_{mn}} e^{-j(k l_{mn} + \varphi_n)+\varphi{'}} \right)^2
\label{eq18}
\end{equation}
\rule{\linewidth}{0.5pt} 
\end{figure*}

Thus, the output power of a MT can be expressed as equation (\ref{eq17}). Similarly to equation (\ref{eq14}),  we can get the power of the electromagnetic wave received by the BS from the MT in (\ref{eq18}).

After the BS receives the electromagnetic wave from the MTs, it reflects the wave back to each MT. However, unlike the MTs, the BS is equipped with power amplification equipment to compensate for the power loss in the system and the workload of the MTs. The power received by the BS is amplified and phase conjugate before re transmitting. Therefore, the signal output from the BS again can be expressed as

\begin{equation}
S_{\text{BS}}^{i,t} = \sqrt{2 \mu_0 f_\text{PA} (P_{\text{BS}}^{i,r})} \cdot \varphi_{\text{BS}}^{i,t}
\end{equation}

The electromagnetic waves between the BS and MT undergo multiple transmissions. Electromagnetic waves with similar phases add up, while waves with different phases cancel each other out, ultimately forming a stable signal between the BS and each MT.

By undergoing multiple electromagnetic wave transmissions between the BS and MT, phase alignment or misalignment leads to either constructive interference (amplification) or destructive interference (diminishing the power). Eventually, all MTs stabilize and maintain a reliable transmission path. The total system transmission efficiency can be defined as

\begin{equation}
\eta = \sum_{i=1}^{I}\frac{P_\text{MT}^{i,r}}{f_\text{PA}(P_{\text{BS}}^{i,r})}.
\end{equation}

\subsection{DOA Estimation}

Through multiple round-trip transmissions, a stable resonance is established between the BS and each MT. At this time, the BS receives both the signal power from the MT and the noise power generated during the iteration process.  These noises are unrelated to the signal and can be estimated for the DOA of the MTs using the MUSIC algorithm \cite{schmidt1986multiple}. 

The covariance matrix obtained by covariance calculation of the signal received by BS is essentially diagonal, and the signal subspace and noise subspace are orthogonal. By using the orthogonality property, the signal subspace and noise subspace can be distinguished \cite{wei2017direction}. This requires searching through the entire steering vector matrix and identifying the eigenvectors that are orthogonal to the noise subspace. The covariance matrix of the signal received by the BS is the expectation of the outer product of the signal defined in equation \ref{eq4} with its conjugate transpose

\begin{equation}
\mathbf{R_X} = \mathbb{E}[\mathbf{X}_\text{BS}^r \cdot {\mathbf{X}_{\text{BS}}^r}^H].
\label{eq21}
\end{equation}

Substituting equation (\ref{eq4}) into (\ref{eq21}), we can get

\begin{align}
\nonumber \mathbf{R_X} &= \mathbb{E}[(\mathbf{A S}_{\text{BS}}^r + \mathbf{P}_\text{N})(\mathbf{A S}_{\text{BS}}^r + \mathbf{P}_\text{N})^H]\\ \nonumber
&= \mathbf{A} \mathbb{E}[\mathbf{S}_{\text{BS}}^r \cdot {\mathbf{S}_{\text{BS}}^r}^H] \mathbf{A}^H + \mathbb{E}[\mathbf{P}_\text{N}\mathbf{P}_\text{N}^H]\\ 
&= \mathbf{A R}_\text{S} \mathbf{A}^H + \mathbf{R}_\text{N},
\end{align}
where, $\mathbf{R}_\text{N} = \sigma^2_\text{N} \textrm{I}$ represents the noise covariance matrix. $\textrm{I}$ is the  identity matrix. When performing eigenvalue decomposition of the covariance matrix, the $K= \sqrt{M}$ eigenvalues are generated, with the larger $I$ eigenvalues corresponding to the signal source and the remaining $K-I$ smaller eigenvalues associated with the noise subspace. The decomposed covariance matrix can be expressed as

\begin{equation}
\mathbf{R}_X = \mathbf{Q}_\text{S} \sum \mathbf{Q}_\text{S}^H + \mathbf{Q}_\text{N} \sum \mathbf{Q}_\text{N}^H,
\end{equation}
where the $\mathbf{Q}_\text{S}$ and $\mathbf{Q}_\text{N}$ represent the basis for the signal subspace and noise subspace, respectively. Since the signal and noise are uncorrelated, we can get

\begin{equation}
\alpha^H(\theta, \phi) \mathbf{Q}_\text{N} \mathbf{Q}_\text{N}^H \alpha(\theta, \phi) = 0.
\label{eq24}
\end{equation}

Thus, by calculating the steering vector $\alpha(\theta, \phi)$ and the noise subspace, the direction of the signal source can be determined. When the angles $\theta$ and $\phi$ approach the true values, equation (\ref{eq24}) tends toward 0, resulting in a peak corresponding to the signal source's DOA in the spectral scan over various angles. The MUSIC spectrum function can be represented as

\begin{equation}
P_{\text{MUSIC}}(\theta, \phi) = \frac{1}{\alpha^H(\theta, \phi) \mathbf{Q}_\text{N} \mathbf{Q}_\text{N}^H \alpha(\theta, \phi)}.
\end{equation}

\section{PERFORMANCE EVALUATION}
In this section, we evaluated the electromagnetic wave transmission and localization performance of the system, we assume that the array configurations of the BS and passive MTs are the same and detailed simulation parameters are shown in the TABLE~\ref{tab:parameter_setting}.

\begin{figure*}
  \centering
        \subfigure[]{
	\includegraphics[width=0.3\linewidth]{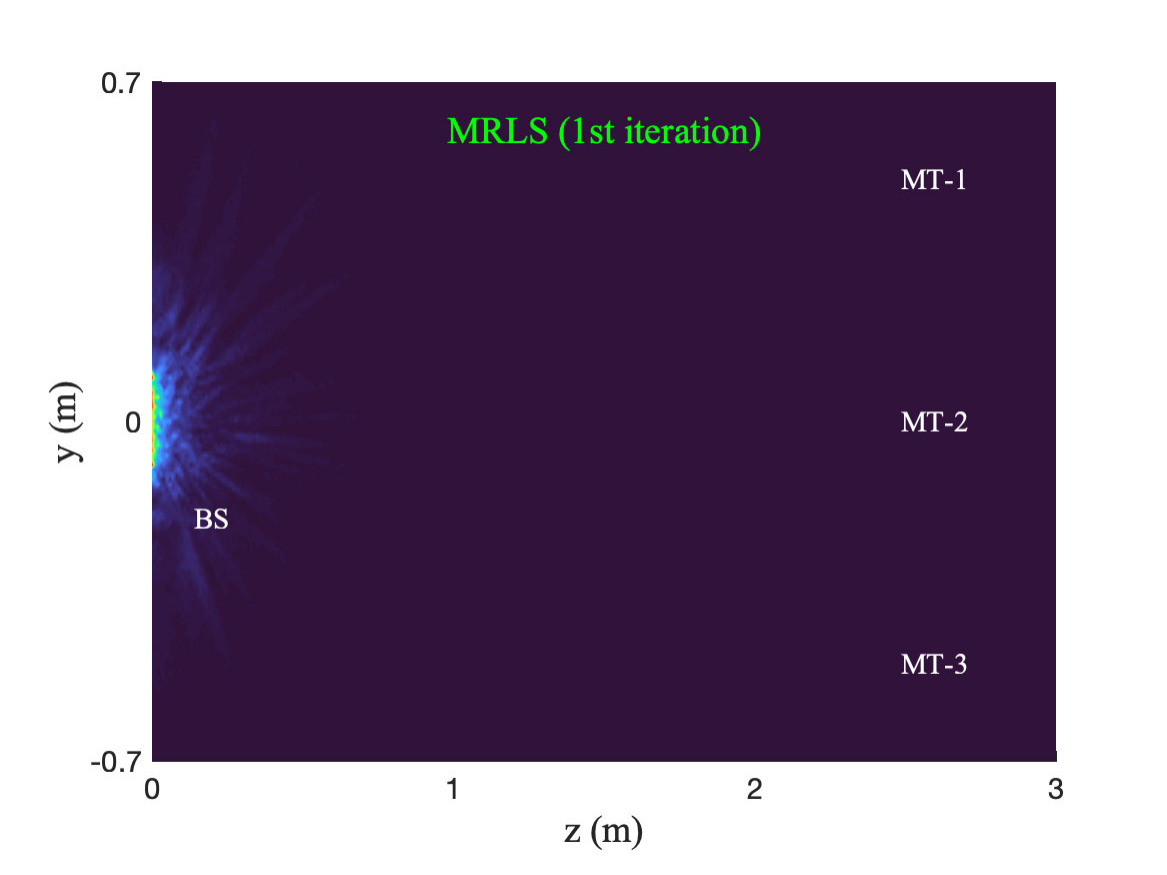}}
        \subfigure[]{
	\includegraphics[width=0.3\linewidth]{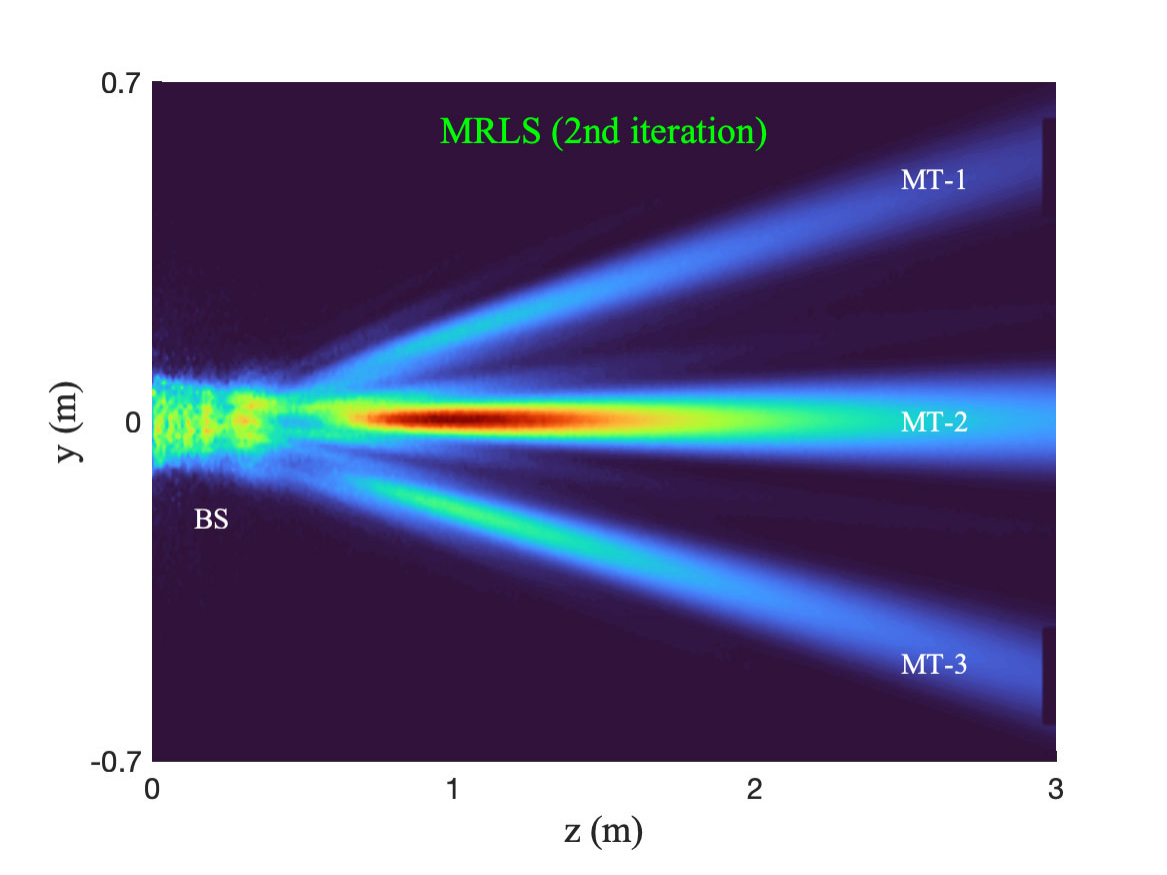}}
        \subfigure[]{
	\includegraphics[width=0.3\linewidth]{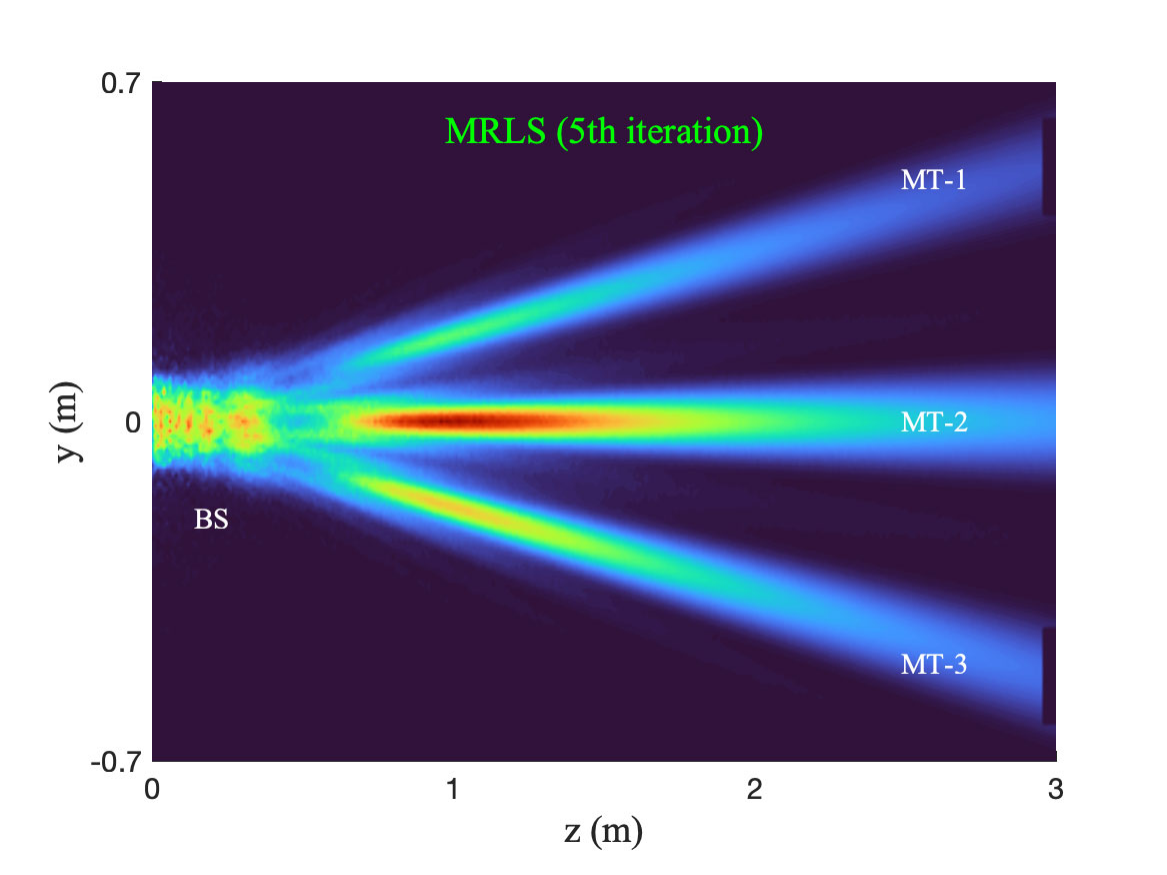}}
        \subfigure[]{
	\includegraphics[width=0.3\linewidth]{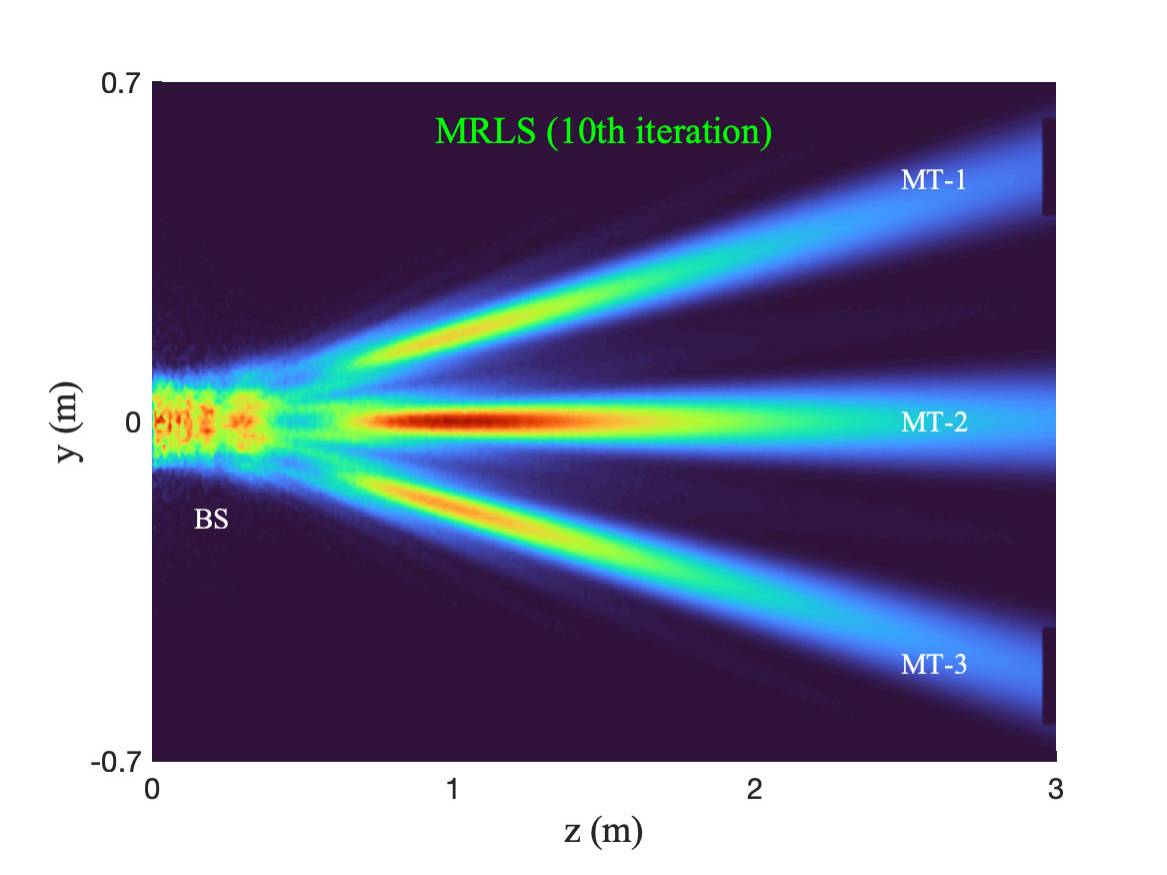}}
        \subfigure[]{
	\includegraphics[width=0.3\linewidth]{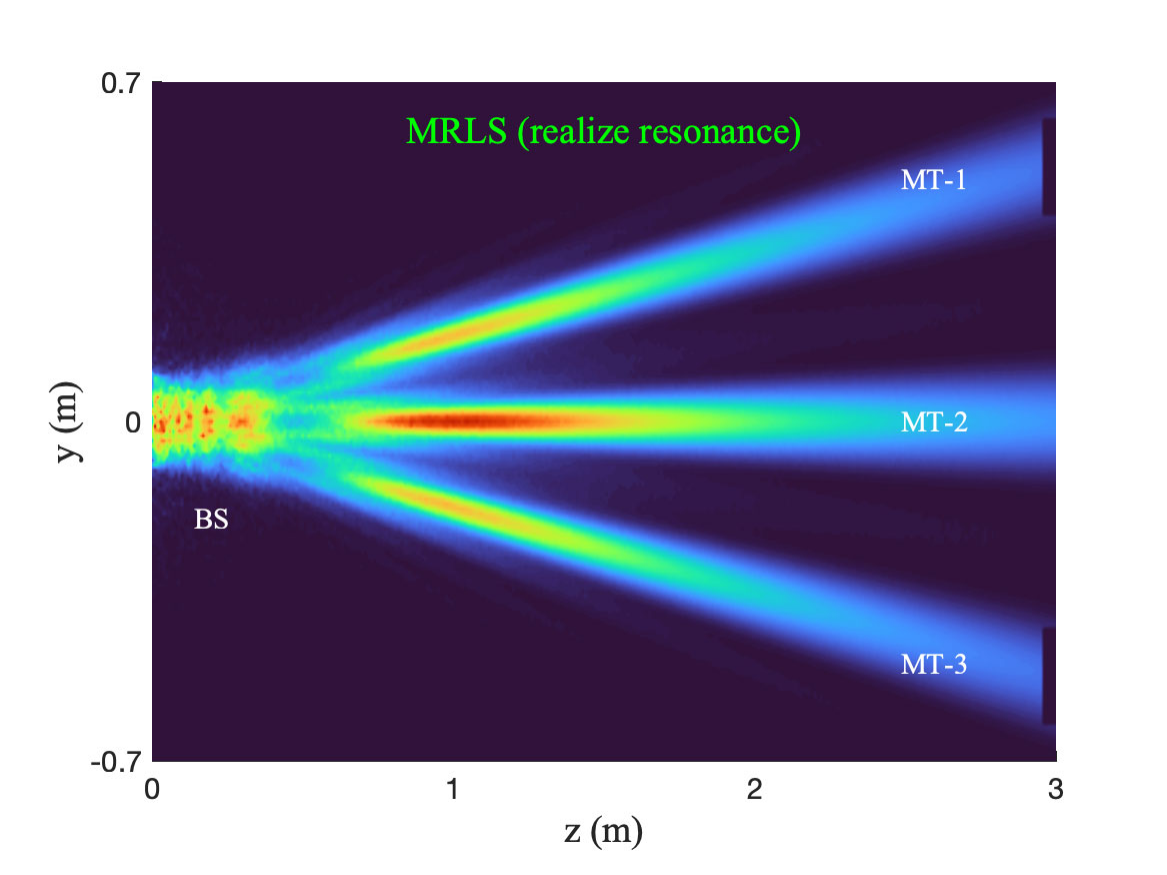}}
        \subfigure[]{
	\includegraphics[width=0.3\linewidth]{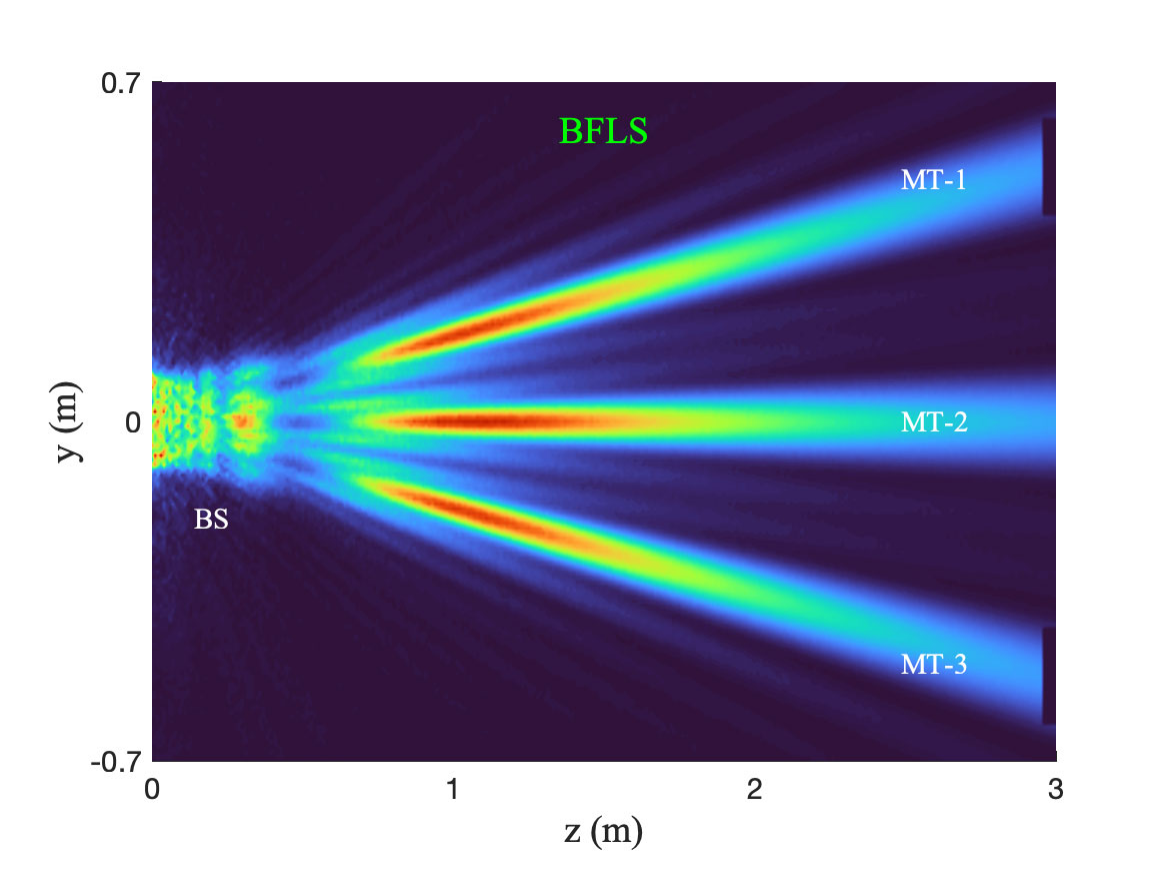}}
\caption{The formation process of MRLS resonance and comparison with BFLS spatial normalized power density. (a) The initial transmission signal of MRLS BS; (b) The power density during the second iteration of MRLS; (c) The power density during the fifth iteration of MRLS; (d) The power density during the tenth iteration of MRLS; (e) The power density during resonance of MRLS; (f) The power density of active beamforming localization system.}
\label{yoz_density}
\end{figure*}

\begin{figure*}
  \centering
        \subfigure[]{
	\includegraphics[width=0.22\linewidth]{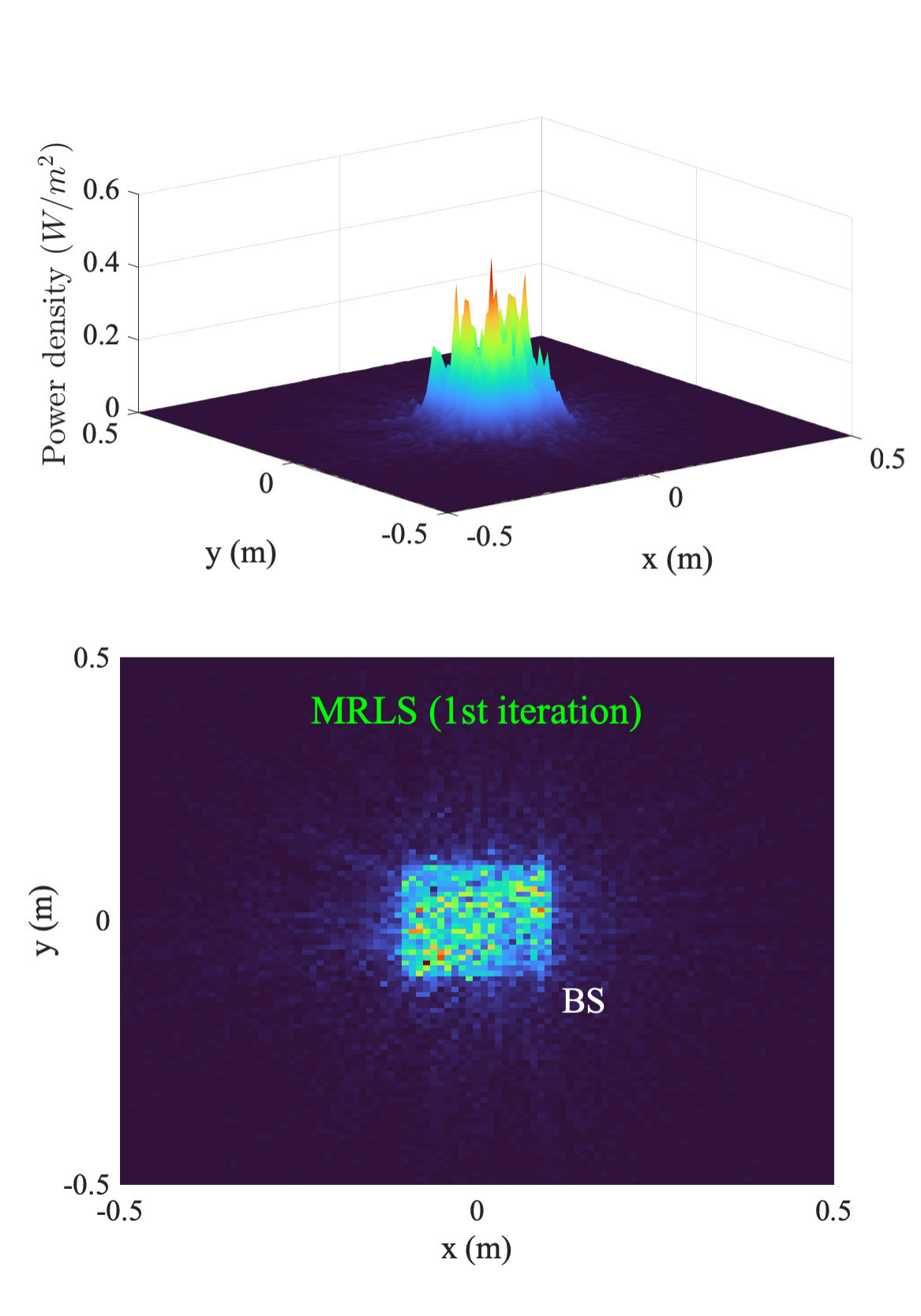}}
        \subfigure[]{
	\includegraphics[width=0.22\linewidth]{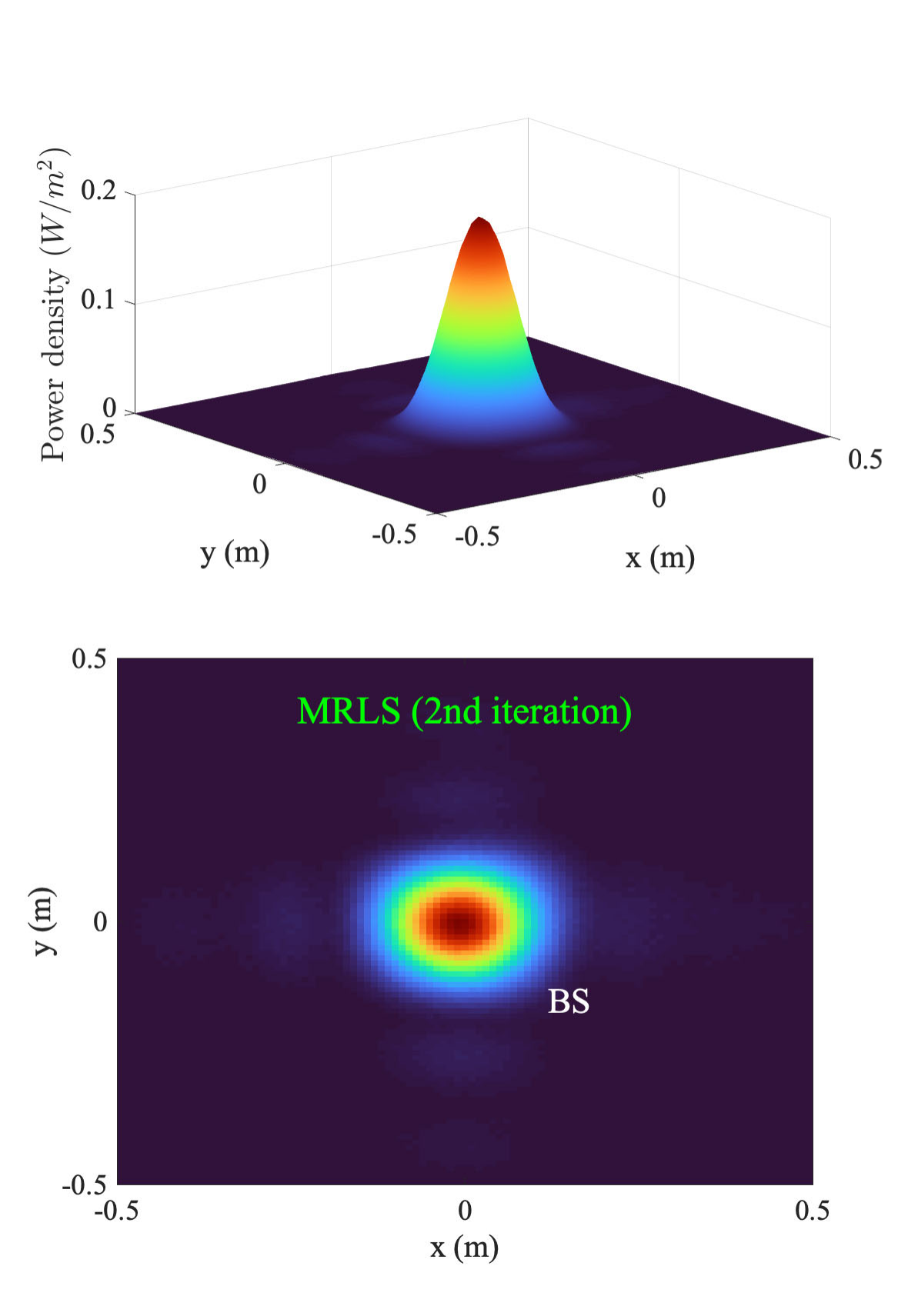}}
        \subfigure[]{
	\includegraphics[width=0.22\linewidth]{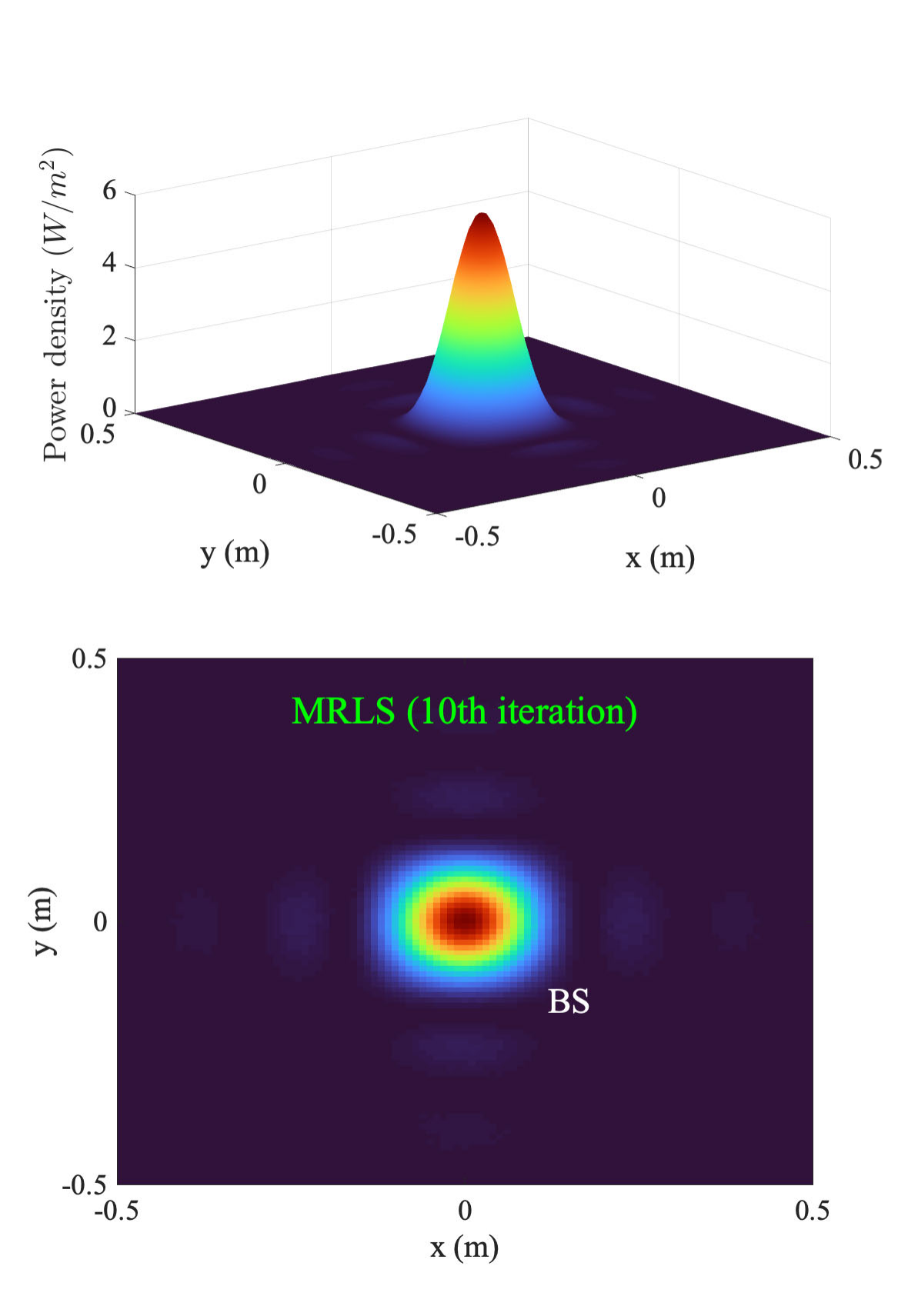}}
        \subfigure[]{
	\includegraphics[width=0.22\linewidth]{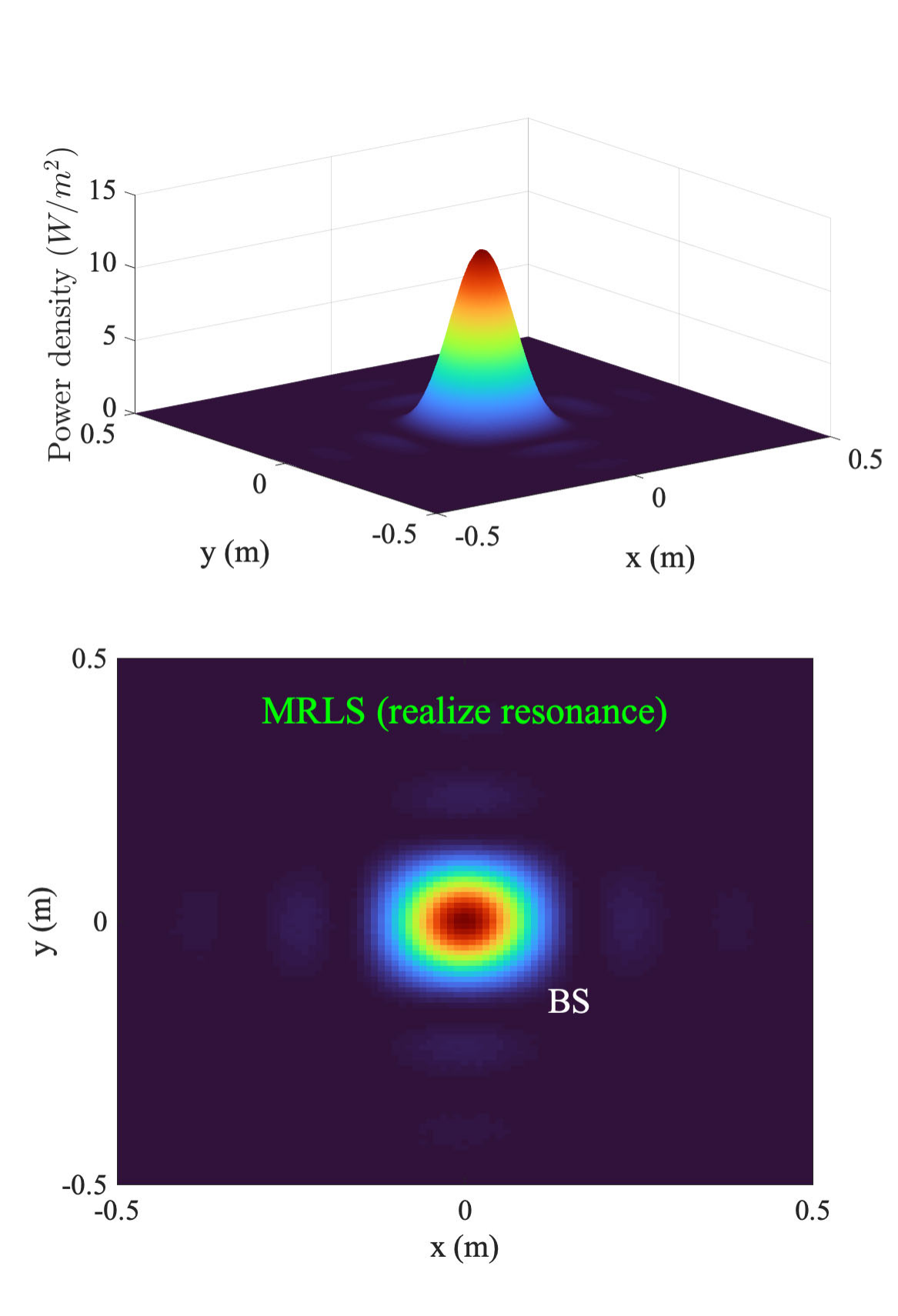}}
        \subfigure[]{
	\includegraphics[width=0.22\linewidth]{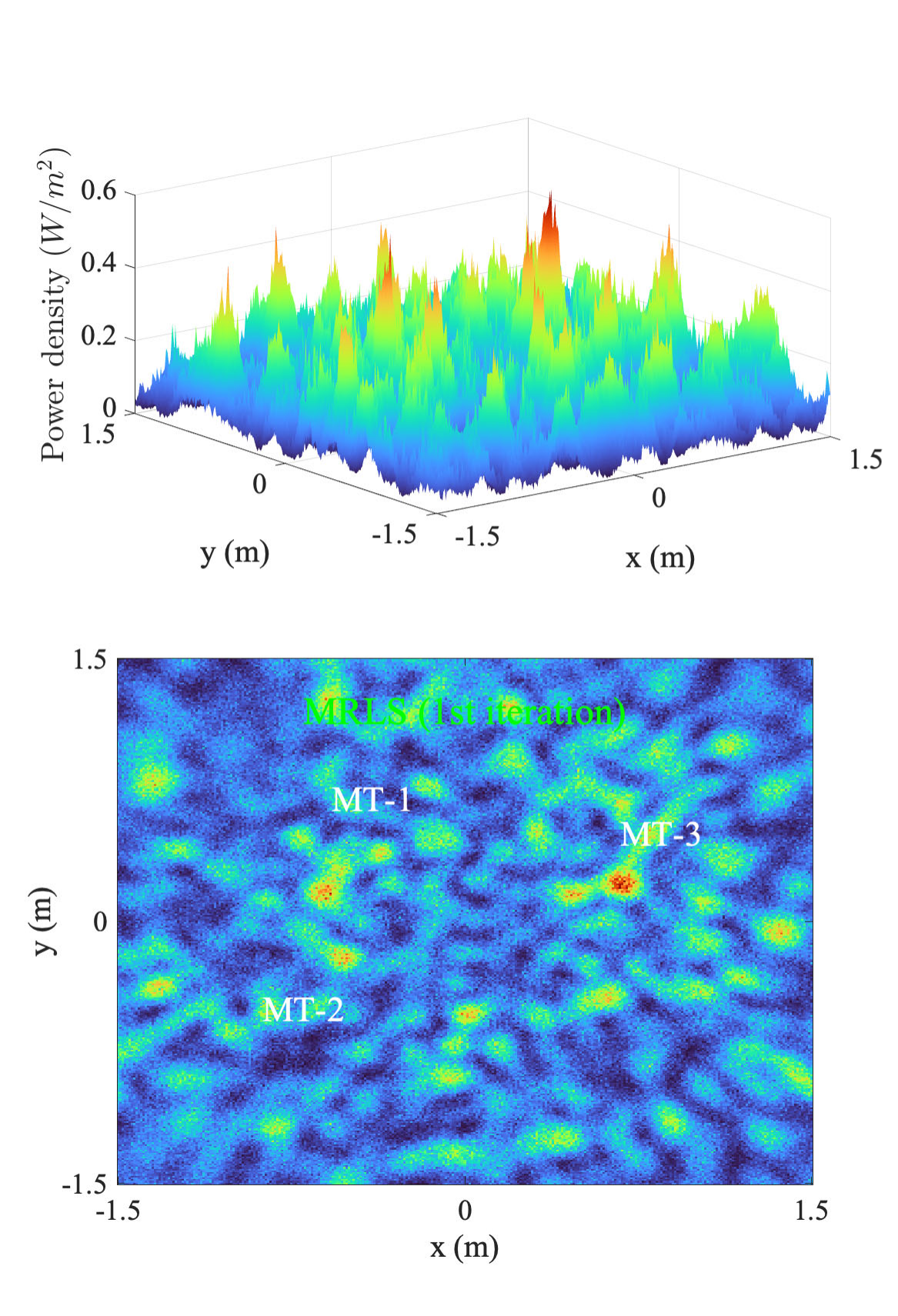}}
        \subfigure[]{
	\includegraphics[width=0.22\linewidth]{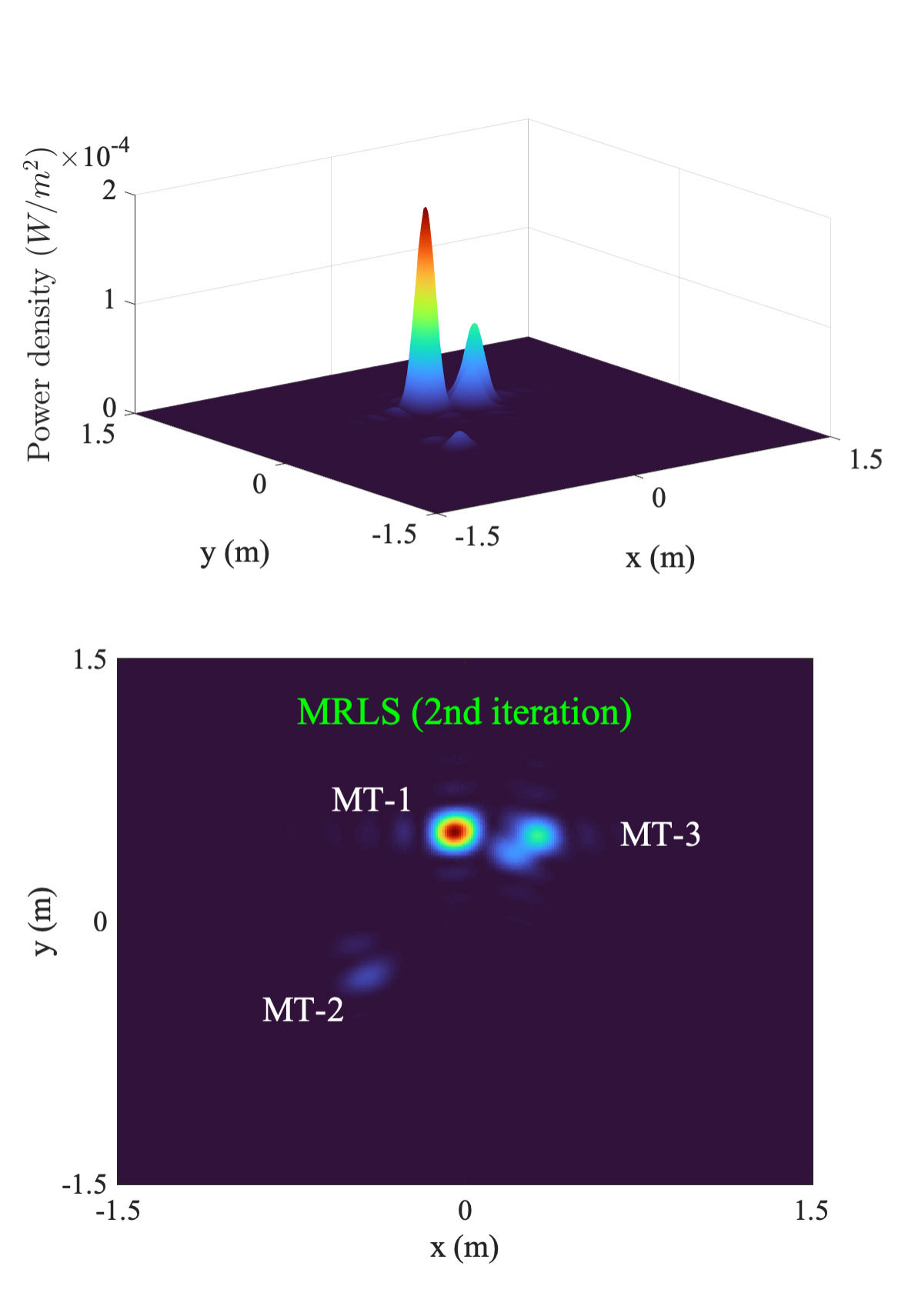}}
        \subfigure[]{
	\includegraphics[width=0.22\linewidth]{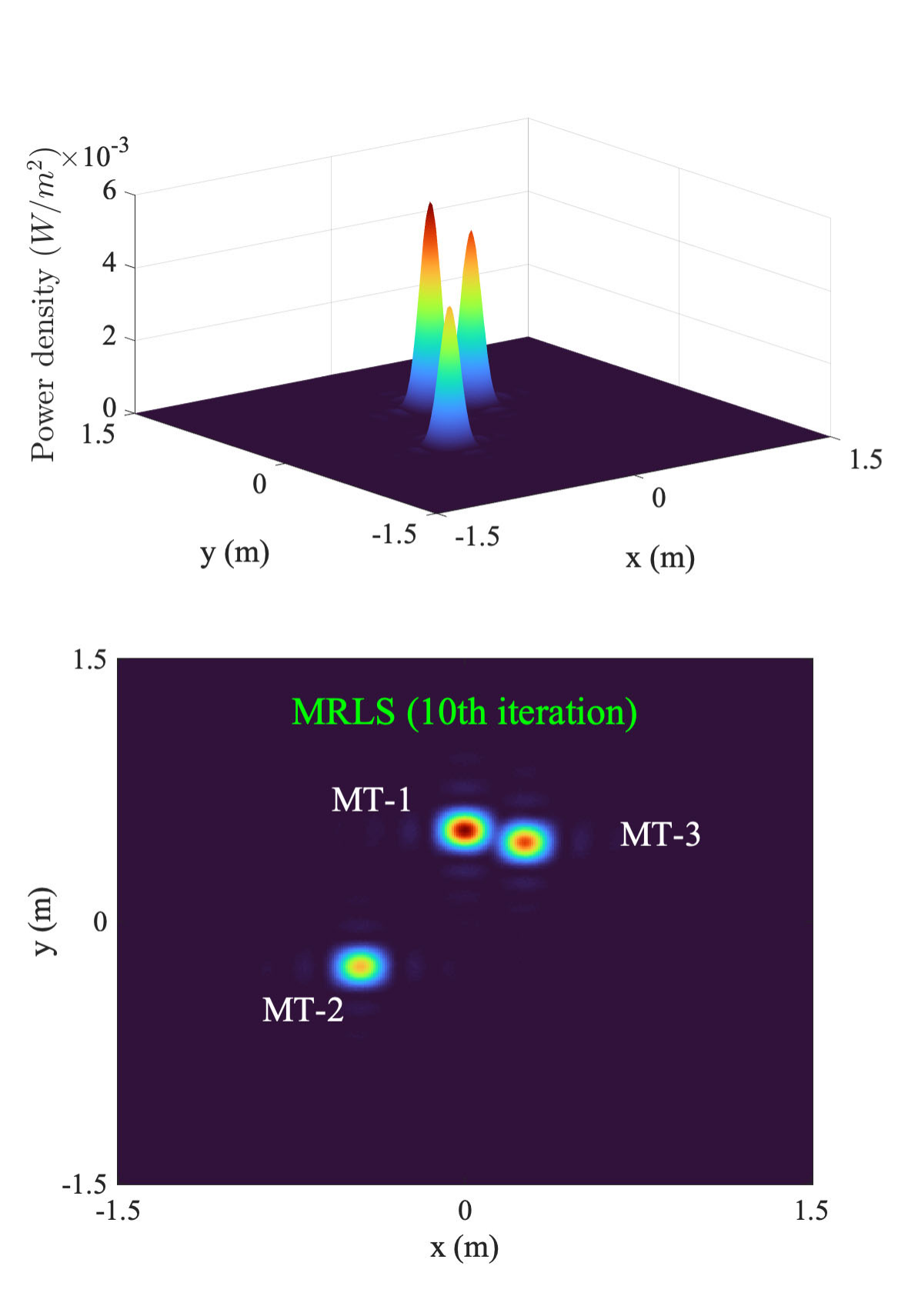}}
        \subfigure[]{
	\includegraphics[width=0.22\linewidth]{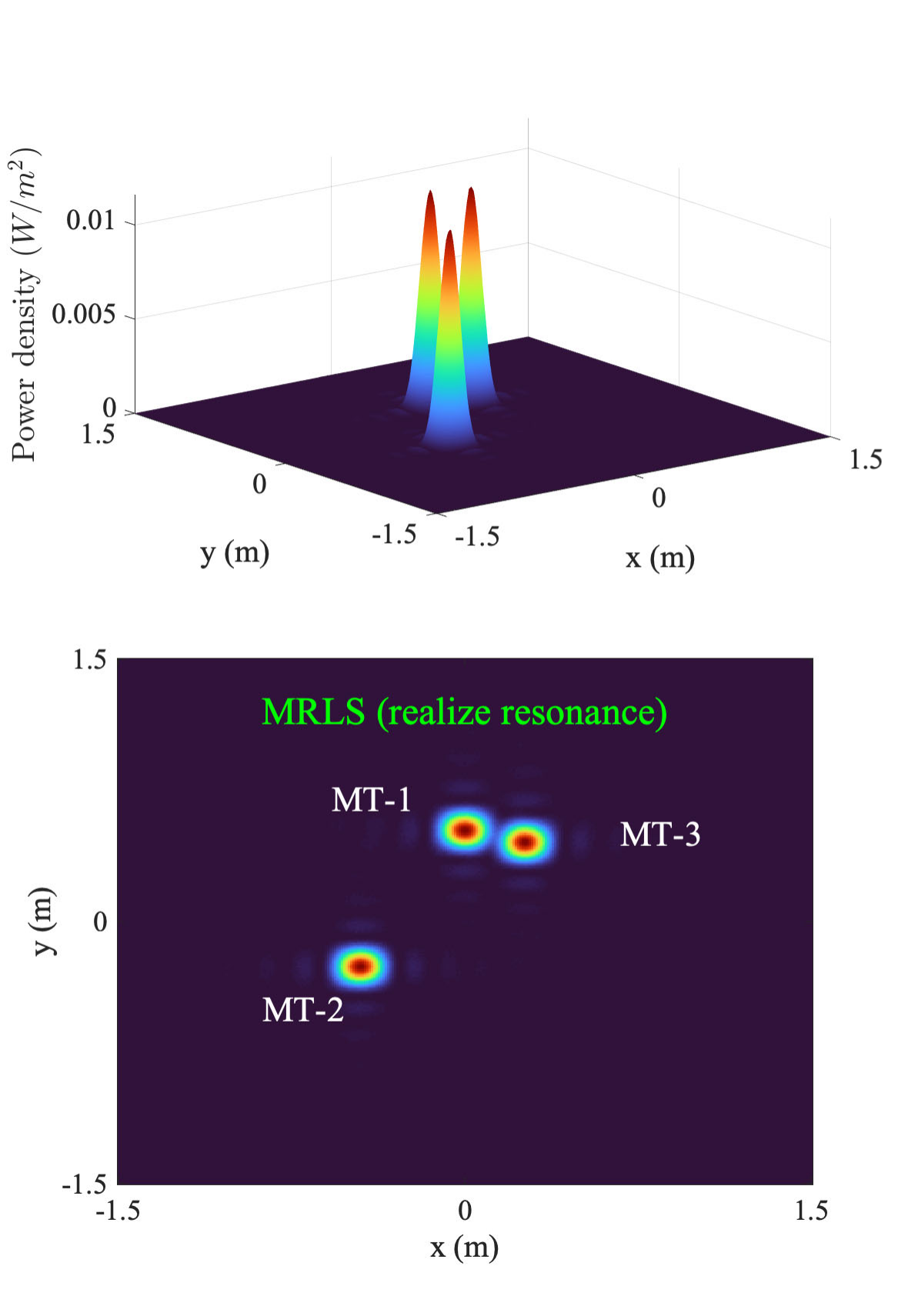}}

\caption{The distribution of system power density during the process of establishing resonance between the BS and three passive MTs simultaneously, (a)-(d) show the BS plane, (e)-(h) show the MT plane.}
\label{density}
\end{figure*}

\begin{figure}[!t]
\centering
\includegraphics[width=0.8\linewidth]{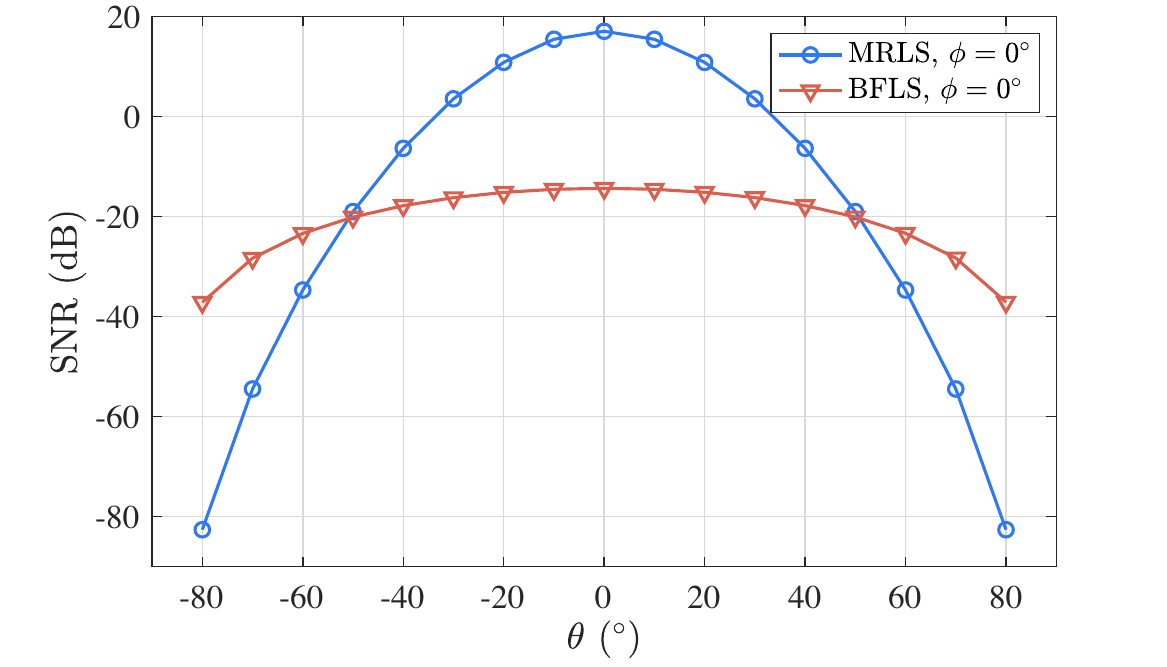}
\caption{The influence of elevation angle on the SNR of MRLS and BFLS.}
\label{SNRvsTheta}
\end{figure}
\subsection{Establishment of Resonance and Performance Analysis}
\begin{table}[h]
    \centering
    \caption{Parameter Setting}
    \begin{tabular}{m{3cm}<{\centering} m{2cm}<{\centering} m{2cm}<{\centering}}
        \toprule 
        \textbf{Parameter} & \textbf{Symbol} & \textbf{Value} \\
        \midrule
        Initial transmission power of BS & $P_\mathrm{BS}^t(0)$ & 1 mW \\
        Distance between BS and each MT & $L$ &  2 m, 3 m\\
        BS center point coordinates & - & (0,0,0) \\
        Maximum gain of antenna \cite{balanis2016antenna}& $G(\theta,\phi)$ & $\leq$4.97 dBi \\     
        MT number & $I$ &  3\\
        Frequency & $f$ & 30 GHz \\
        Wavelength & $\lambda$ & 1 cm \\
        Element spacing & $\lambda/2$ & 0.5 cm \\
                Reflection ratio & $\beta$ & 0.4\% \\
        Array size & $M, N$ & 40$\times$40, 30$\times$30\\
        Amplifier gain \cite{devices2023}& $G_\mathrm{PA}$ & $\leq$24 dB \\
        Variance of phase noise & $\sigma_{\varphi'}^2$ & 0.3 rad$^2$ \\
        \bottomrule
    \end{tabular}
    \label{tab:parameter_setting}
\end{table}

Figure~\ref{yoz_density} shows the normalized power density in the $yoz$-plane during the process of resonance formation between the BS and three passive MTs, compared with the active beamforming localization system (BFLS) in Fig.~\ref{yoz_density}(f). Figs~\ref{yoz_density}(a), (b), (c), and (d) at iterations 1, 2, 5, and 10 respectively, and Fig.~\ref{yoz_density}(e) is the final resonance effect achieved. The configurations between all MTs and the BS are identical, with each comprising 40$\times$40 elements and a transmission distance of 3 m. The elevation and azimuth angles between each MT and the BS are as follows: MT-1 (10$^\circ$, 0$^\circ$), MT-2 (0$^\circ$, 0$^\circ$), and MT-3 (-10$^\circ$, 0$^\circ$). Initially, the BS emits omnidirectional electromagnetic waves with a power of 1 mW into the space. Each passive MT, upon receiving electromagnetic waves from the BS, reflects back a fixed proportion of the electromagnetic waves to the BS using RDA. At this stage, the transmission efficiency is very low. The BS then applies phase conjugation to the electromagnetic waves received from each MT to enable directional transmission. Additionally, the BS amplifies the received power using a power amplifier and transmits the amplified power, marking the start of the second iteration.

As the power amplifier continues to operate in multiple iterations, it compensates for transmission losses and provides the necessary power output for all MTs. Eventually, the system reaches a stable state where the transmission efficiency is maximized. The system is considered to have reached steady-state resonance when the power difference received by the BS in two consecutive iterations is less than 0.001\% of the received power in the previous iteration.

During the resonance process in MRLS, it can be observed that the power density of the resonant beam between MT-2 and the BS is the highest. This is because the antenna gain of MT-2 is at its maximum in this configuration. Although MT and \mbox{MT-3} have the same configuration and absolute elevation angles relative to the BS, their power densities are not consistently identical due to the presence of phase noise. In the early stages of iteration, random phase noise affects the input and output power of each element. When the system achieves resonance, the spatial power densities of the two MTs become consistent, and the power transmission efficiency reaches its maximum, with the impact of phase noise becoming negligible.

Furthermore, as shown in Figs.~\ref{yoz_density}(e) and (f), the MRLS system in resonance demonstrates significant advantages over the BFLS system. MRLS not only achieves automatic beam alignment but also exhibits reduced sidelobes and stronger directionality.

Figure~\ref{density} further illustrates the signal power density distribution in the $xoy$-plane during the resonance formation process between the BS and three passive MTs. 
Figs.~\ref{density}(a), (b), (c), and (d) are the distribution of power density in BS plane, Figs.~\ref{density}(e), (f), (g), and (h) are the distribution of power density in MT plane.
The elevation and azimuth angles of each MT are: \mbox{MT-1} (10$^\circ$, 30$^\circ$), MT-2 (-10$^\circ$, 30$^\circ$), and MT-3 (10$^\circ$, 60$^\circ$). The array configurations are all 40$\times$40, and the transmission distances are all 3 m. The upper part of each figure shows the actual power density distribution in units of watts per square meter, while the lower part shows the normalized power density in the $xoy$-plane. It can be clearly seen that at the initial moment, the electromagnetic waves radiated by the BS are randomly omnidirectional due to the random phase of each array element, resulting in a relatively chaotic distribution of electromagnetic waves in the MT plane, with low power density and a wide distribution range. From the second iteration onward, the distribution of electromagnetic waves begins to concentrate, especially around the three MTs, where the power density starts to increase significantly, indicating the initial formation of beam concentration. With further iterations, the power density distribution on both the BS plane and the MT planes becomes more concentrated, and the power density values continue to increase, ultimately achieving resonance at the maximum power density. This further verifies that the proposed system not only features automatic beam alignment and highly concentrated energy but also can achieve stable resonance with multi-target in space simultaneously. It is worth noting that the power density on the MT planes after reaching resonance is not the same, with MT-3 having significantly lower power density than MT-1 and MT-2, which is due to the effect of elevation angle on system transmission efficiency. Further analysis on the impact of elevation angle on system performance is subsequently carried out.

Figure~\ref{SNRvsTheta} depicts the trend of SNR for the active BFLS and MRLS as elevation $\theta$ changes. Here, we set the azimuth angle $\phi=0^\circ$, and both systems use a 40$\times$40 array, with the BS subject to 0.03 mW of noise power interference. It can be seen that in the range from -50$^\circ$ to 50$^\circ$, the SNR of MRLS is significantly higher than that of BFLS. This is because, within this elevation range, the signal power received by the BS in the MRLS continues to increase as resonance is formed, meaning that the amplified power and antenna gain can compensate for the propagation losses of the signal.

However, beyond (-50$^\circ$, 50$^\circ$), the SNR of BFLS is higher. This is because the elevation severely affects the antenna gain of the system, and as the elevation increases, the antenna gain decreases. When the angle becomes large enough, low signal gain and power amplifiers are no longer sufficient to compensate for transmission losses. Therefore, with increasing MRLS iterations, the power received by the BS continues to decrease, leading to a rapid decline in SNR as noise power remains constant. On the other hand, the BFLS only requires one transmission, so even as the angle increases, the power received by the BS decreases more smoothly. Therefore, it can be concluded that the optimal working range of MRLS is from -50$^\circ$ to 50$^\circ$, and thus in subsequent simulations, we limit the elevation angle range to 0$^\circ$ to 50$^\circ$.

\begin{figure}[!t]
\centering
\includegraphics[width=0.9\linewidth]{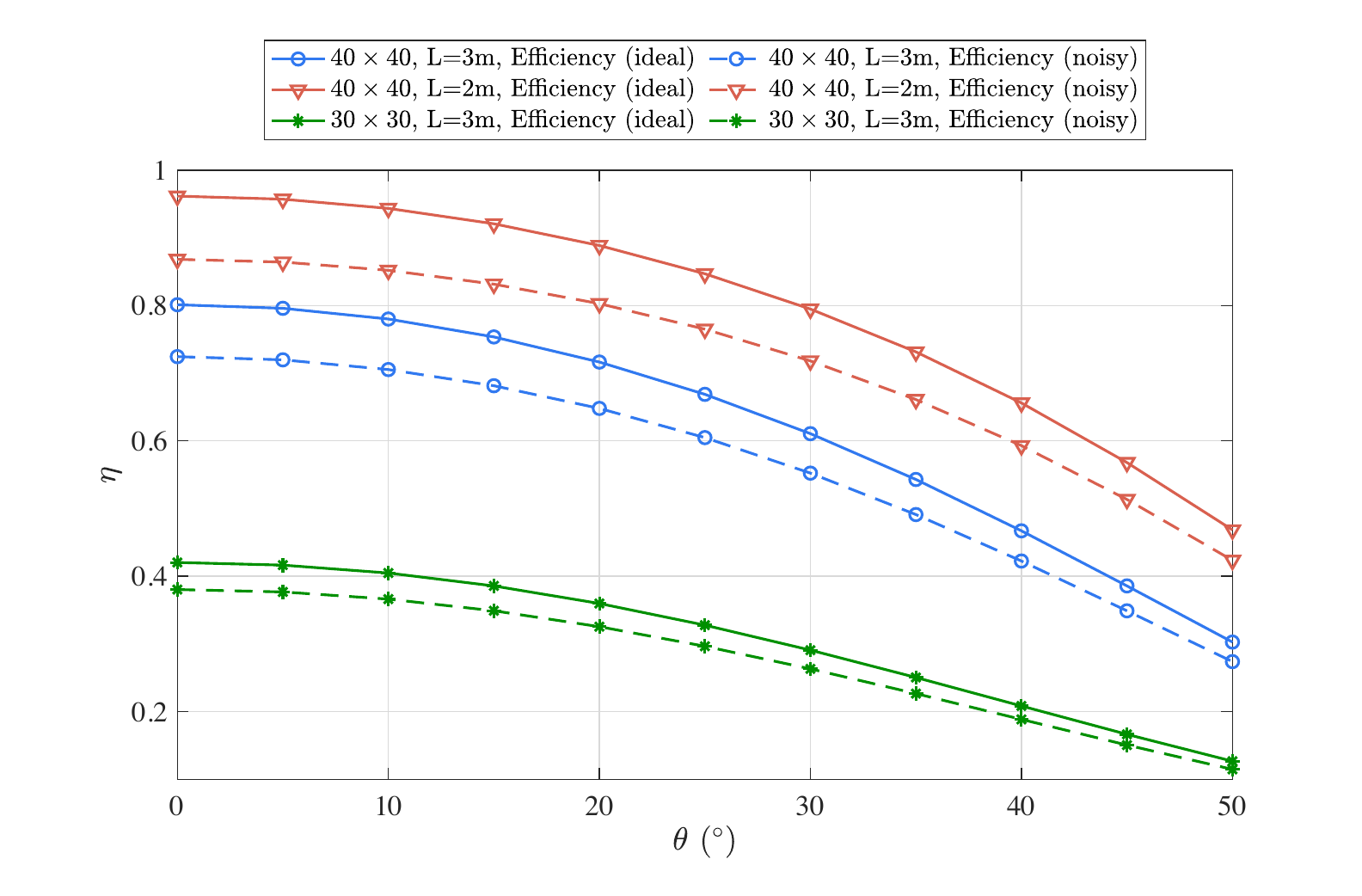}
\caption{The influence of elevation angle on the power transmission efficiency of MRLS under different distances and array sizes.}
\label{ETAvsTheta}
\end{figure}

Figure.~\ref{ETAvsTheta} illustrates the relationship between transmission efficiency and elevation angle at different localization distances and array sizes. The ideal state here refers to the absence of phase noise. To make the results solid, we conducted 100 simulations and took the average. It can be seen that as the elevation angle increases, the transmission efficiency of MRLS decreases. This is because the change in elevation angle will cause the effective receiving area of the RDA antenna to decrease, making it unable to operate at its optimal state. In addition, the increase in distance and the decrease in array size will exacerbate this change, which is due to the deterioration of communication conditions.

\begin{figure}[!t]
\centering
	\includegraphics[width=0.9\linewidth]{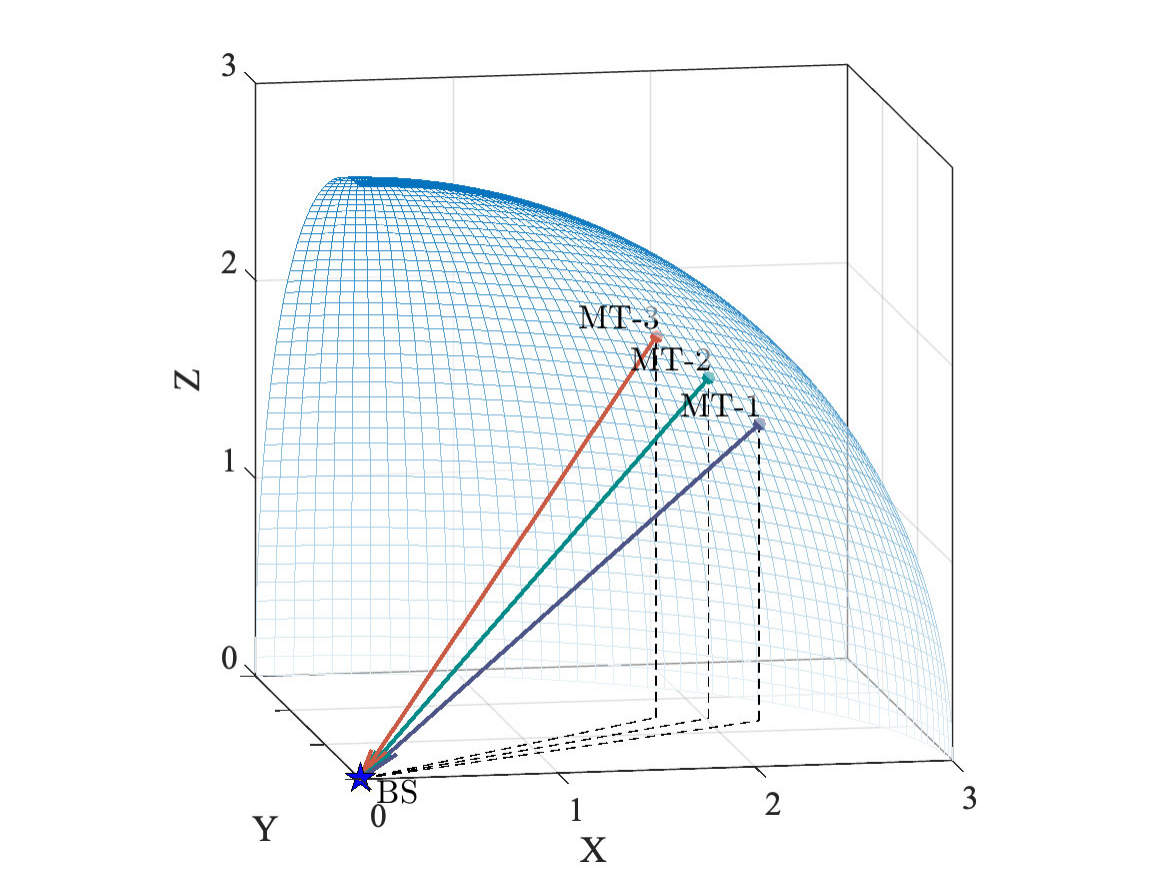}
\caption{The system model for multi-target localization in space by a BS.}
\label{MUSICmodel}
\end{figure}

\begin{figure*}[!t]
\centering
 \subfigure[]{
	\includegraphics[width=0.3\linewidth]{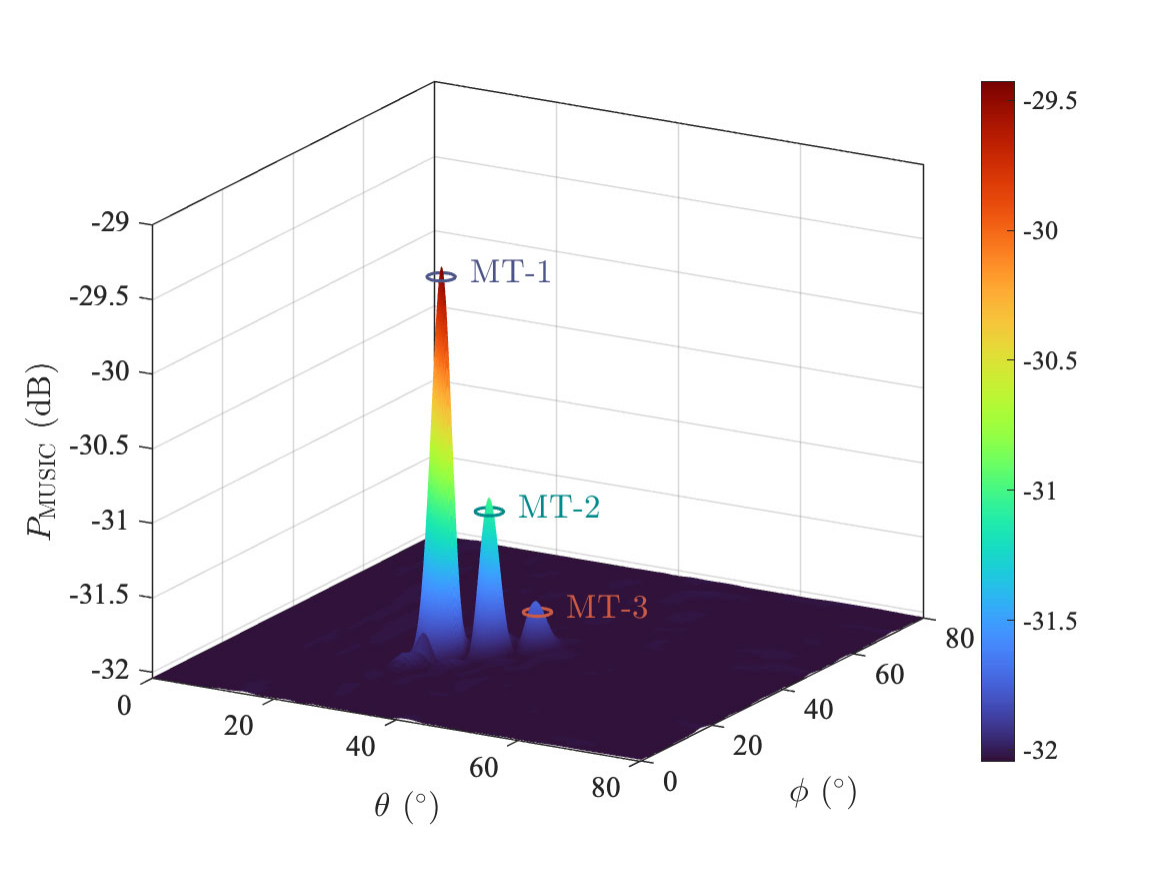}}
 \subfigure[]{
	\includegraphics[width=0.3\linewidth]{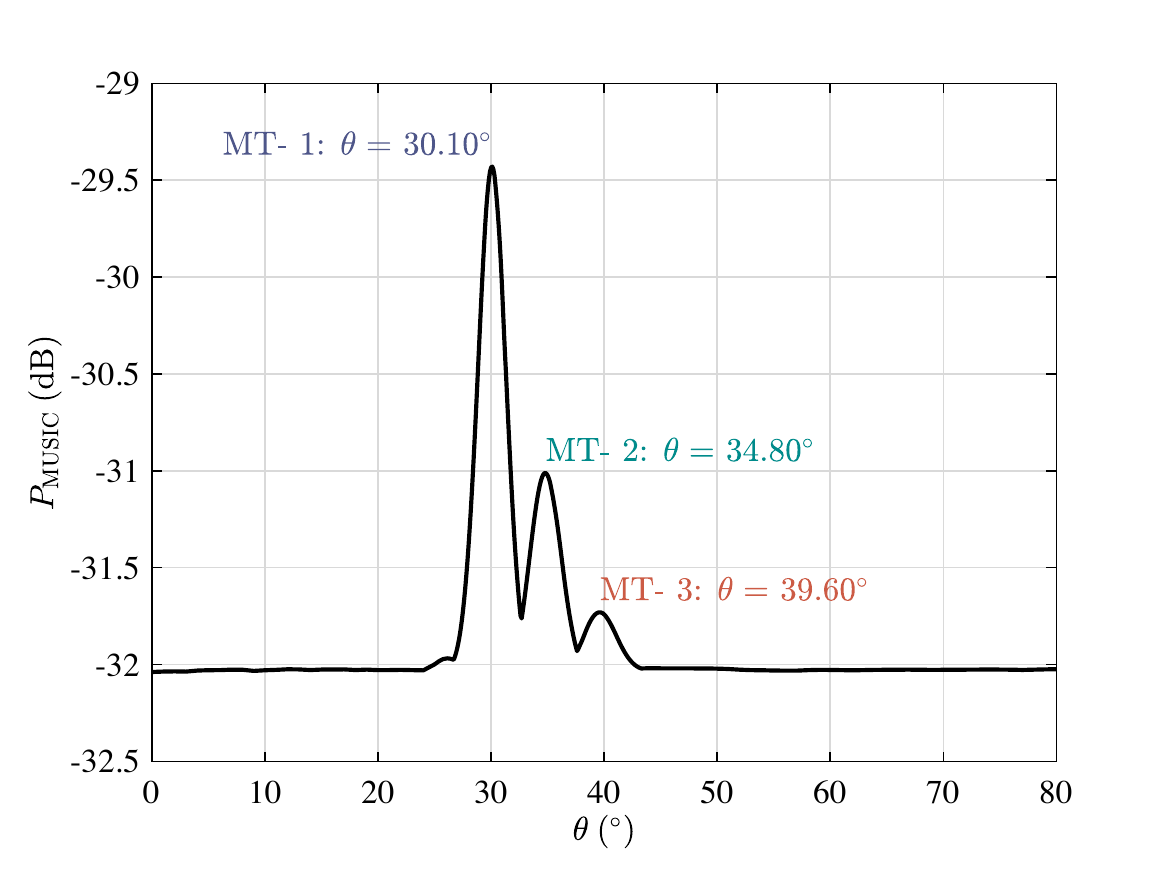}}
 \subfigure[]{
	\includegraphics[width=0.3\linewidth]{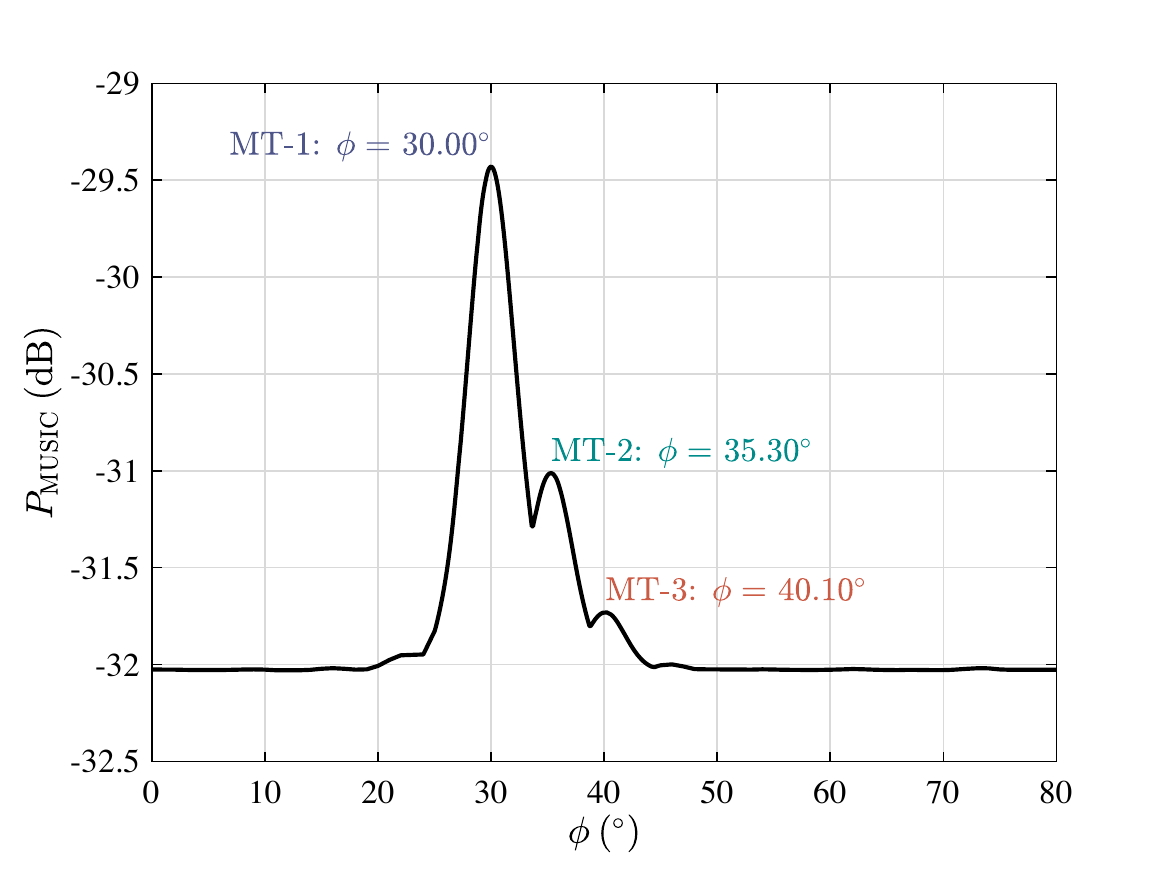}}
\subfigure[]{
	\includegraphics[width=0.3\linewidth]{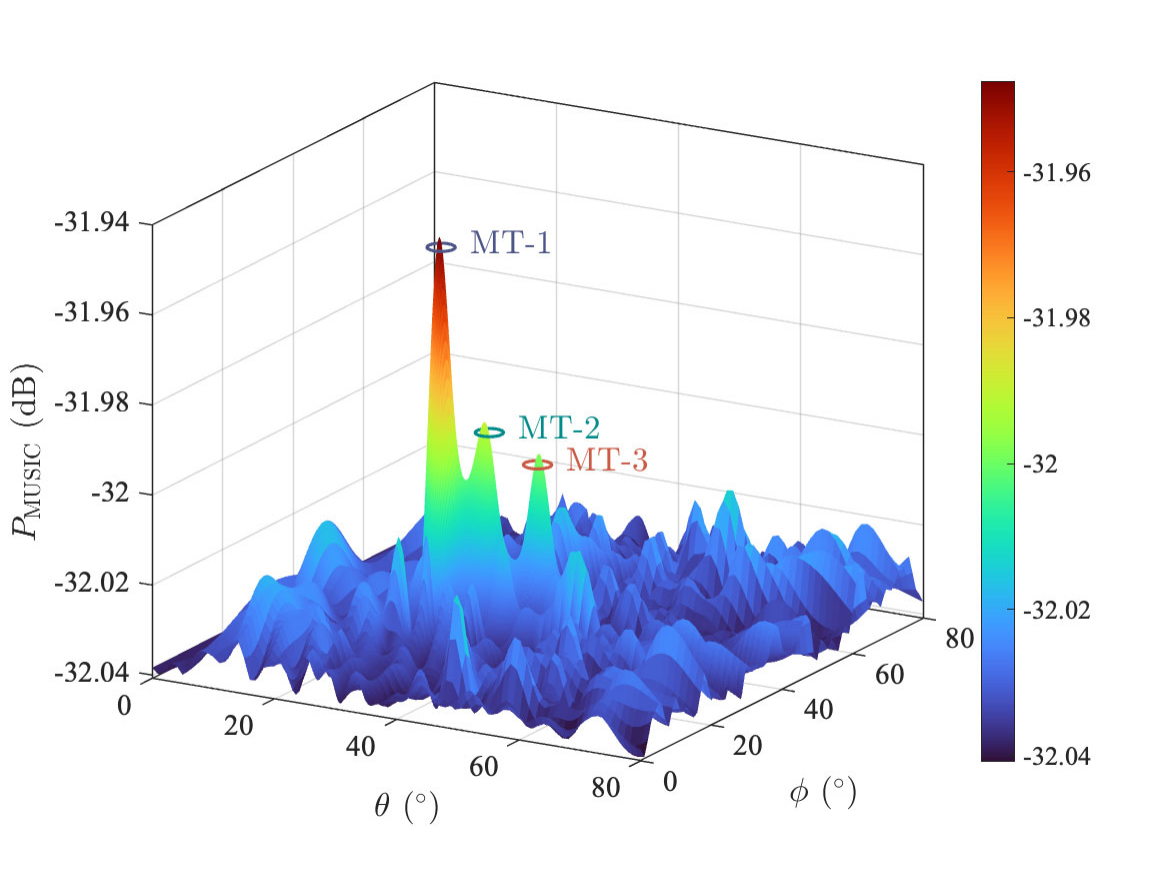}}
 \subfigure[]{
	\includegraphics[width=0.3\linewidth]{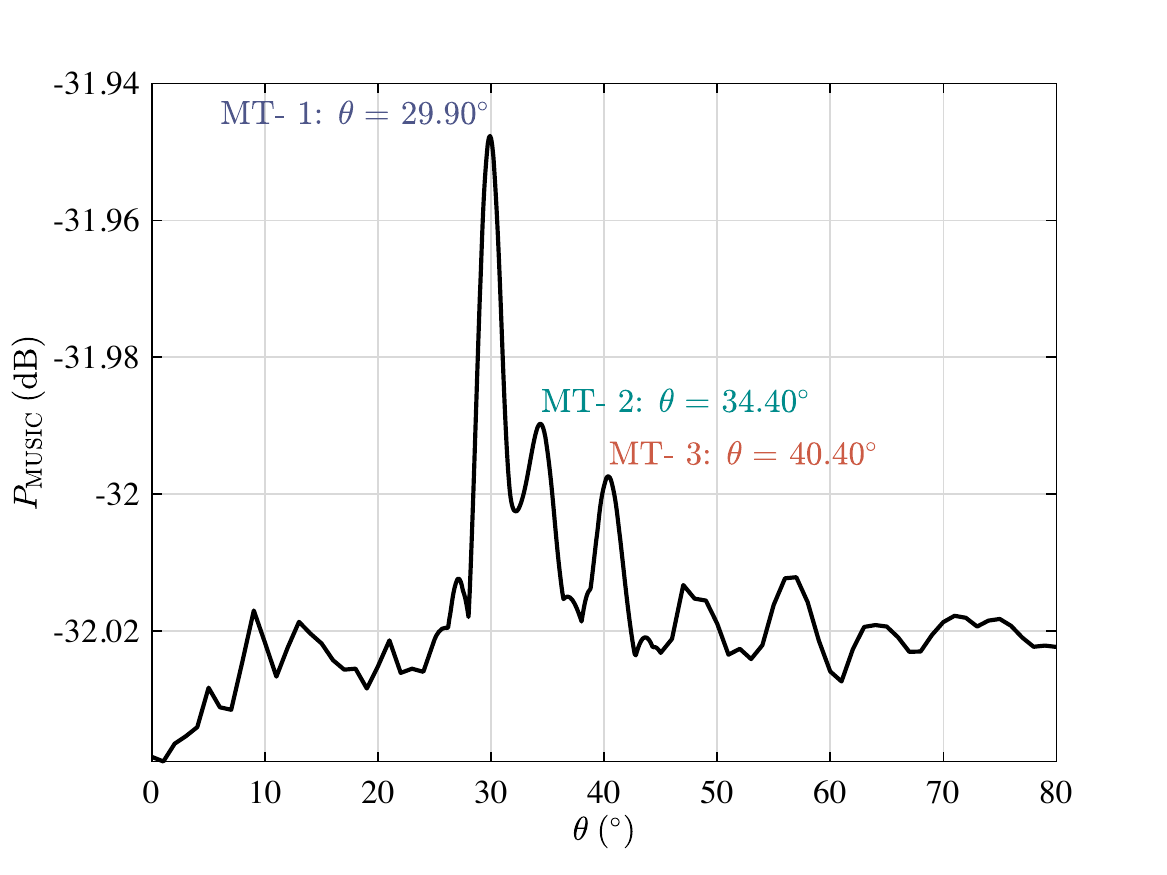}}
 \subfigure[]{
	\includegraphics[width=0.3\linewidth]{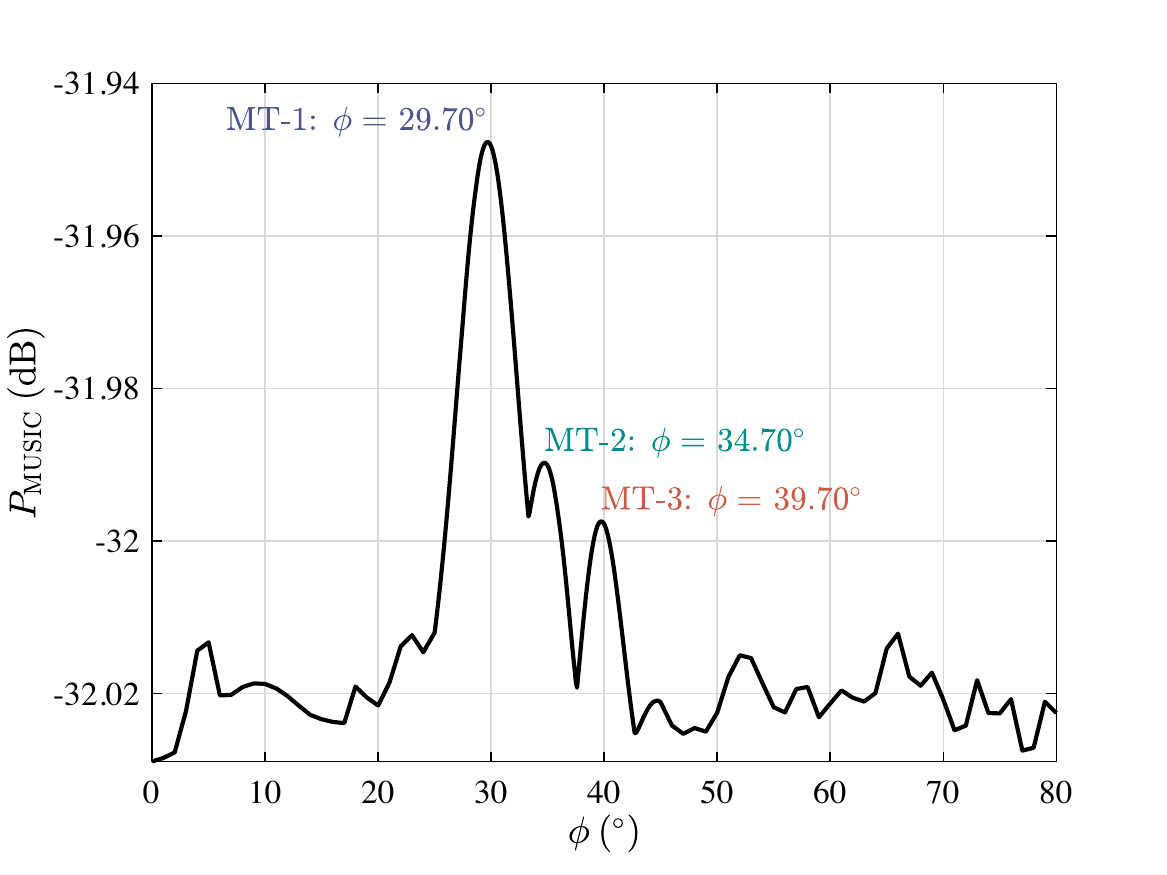}}
\caption{The results of estimating DOA in multi-target scenarios using MUSIC algorithm: (a) Three-dimensional MUSIC spectrogram of MRLS; (b) Elevation estimation results of MRLS; (c) Azimuth estimation results of MRLS; (d) Three-dimensional MUSIC spectrogram of BFLS; (e) Elevation estimation results of BFLS; (f) Azimuth estimation results of BFLS.}
\label{MUSIC}
\end{figure*}

\subsection{DOA Estimation Accuracy Analysis}
In order to locate multiple MTs simultaneously, it is necessary to track and estimate the feature vectors associated with each MT simultaneously. In this subsection, we evaluate the localization performance of MRLS and BFLS in a multi-target scenario using the MUSIC algorithm. The true elevation and azimuth angles for each MT are as follows: MT-1 (30$^\circ$, 30$^\circ$), MT-2 (35$^\circ$, 35$^\circ$), and MT-3 (40$^\circ$, 40$^\circ$), with each MT at a distance of 3 m from the BS, as shown in Fig.~\ref{MUSICmodel}, and the BS is also affected by a noise power of 0.03 mW while receiving signal power.

Figure~\ref{MUSIC} shows the DOA estimation results for MRLS and BFLS using the MUSIC algorithm. Figs.~\ref{MUSIC}(a), (b), and (c) respectively show the three-dimensional MUSIC spectrogram, elevation angle estimation results, and azimuth angle estimation results of MRLS. Figs.~\ref{MUSIC}(d), (e), and (f) correspond to BFLS. It can be seen that in both systems, the larger the elevation angle, the greater the estimation error and the lower the spectrum peak. This is because an increase in elevation angle leads to a decrease in SNR, and simultaneously reduces transmission efficiency and antenna gain, thereby lowering the signal power received by the BS. It is also evident that the DOA estimation accuracy of MRLS is significantly better than that of BFLS, and the entire graph appears smoother, with higher spectrum peaks. This advantage can be attributed to the self-alignment and energy concentration features of MRLS, which ensure that even under passive conditions, once the system achieves resonance, the signal power received by the BS remains much higher than the noise power, despite the presence of the same noise level.  In contrast, the non-smooth spectrogram of BFLS may lead to spectral peaks from non-main directions being higher than those from the actual direction, severely affecting localization accuracy.

\begin{figure}[!t]
\centering
\includegraphics[width=0.9\linewidth]{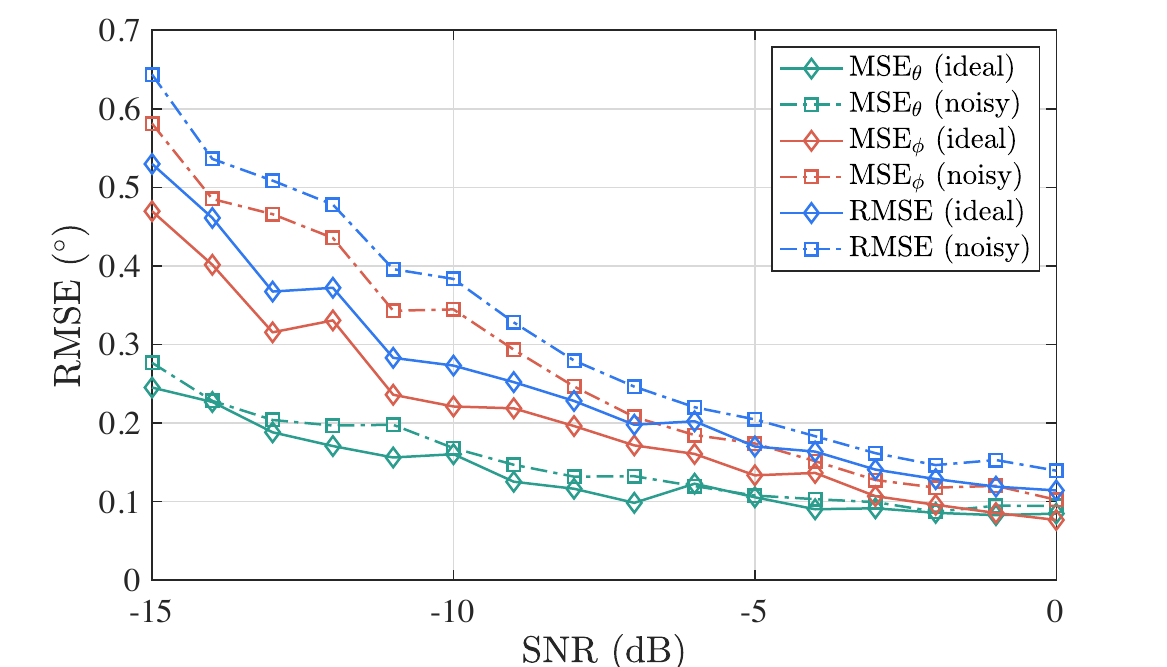}
\caption{The RMSE of MRLS in DOA estimation under different SNR conditions.}
\label{snrVSrmse}
\end{figure}

As shown in Fig.~\ref{snrVSrmse}, we present the impact of different SNR levels on the DOA estimation of the proposed MRLS. We quantify the DOA estimation performance employing the root mean square error (RMSE) which can be expressed by 
\begin{equation}
\mathrm{RMSE}=\sqrt{\mathrm{MSE}_{\theta_i}+\mathrm{MSE}_{\phi_i}},
\end{equation}
where the $\mathrm{MSE}_{\theta_i}=E[(\theta_i-\hat{\theta_i})^2]$ is the mean square error (MSE) of the elevation, and the $\mathrm{MSE}_{\phi_i}=E[(\phi_i-\hat{\phi_i})^2]$ is the MSE of the azimuth, the $\theta_i=30^\circ$ and $\phi_i=15^\circ$ are true values of $i$-th angle, the $\hat{\theta_i}$ and $\hat{\phi_i}$ are the estimation of the $i$-th angle. We conducted 200 Monte Carlo simulation experiments and observed that increasing the SNR reduces the MSE of elevation and azimuth angles, as well as the RMSE of the entire system's DOA estimation, thereby improving the system's localization accuracy. Moreover, the presence of phase noise can affect the input phase of the MUSIC algorithm, leading to an increase in DOA estimation error.

\begin{figure}[!t]
\centering
\includegraphics[width=0.9\linewidth]{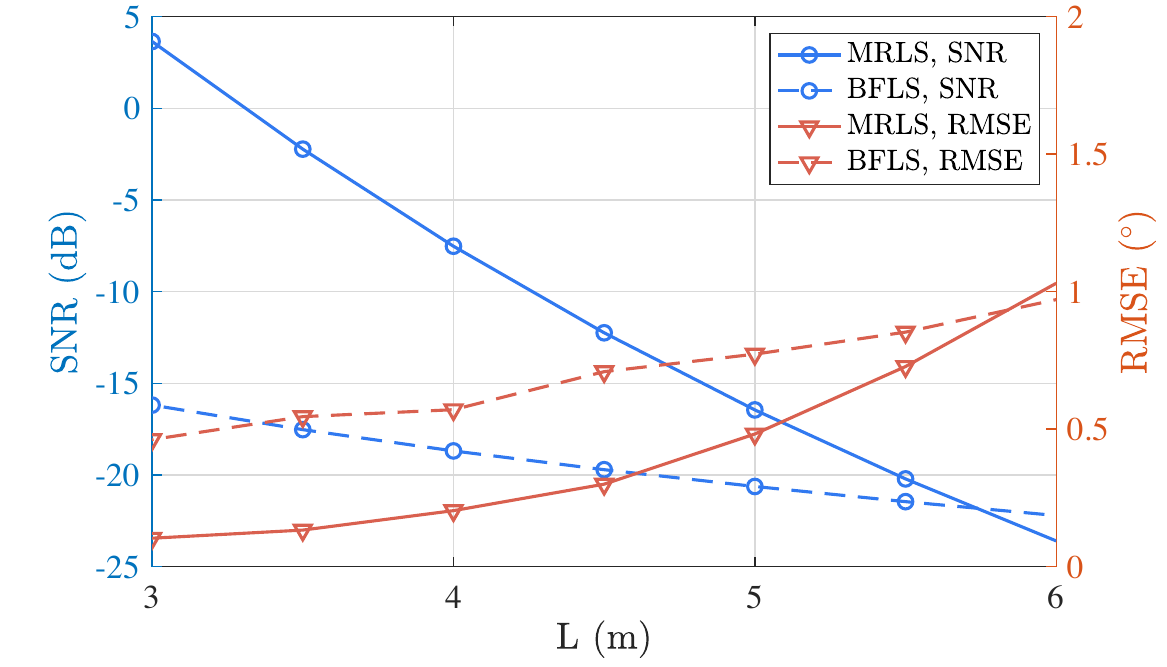}
\caption{The SNR and RMSE of MRLS and BFLS in DOA estimation under different distance between BS and MT.}
\label{lVSrmse}
\end{figure}

In Fig.~\ref{lVSrmse}, we compare the SNR and DOA estimation RMSE of MRLS and BFLS at different distances. Here, we assume an elevation angle of 30$^\circ$ and an azimuth angle of 15 $^\circ$ between the MT and BS, and conducted 100 simulation experiments. It can be observed that the SNR of MRLS is initially significantly higher than that of BFLS, but as the distance increases, the SNR difference between the two systems gradually diminishes. Finally, at a distance of 6 m, the SNR of BFLS becomes higher than that of MRLS. This is because, in MRLS, the MT is passive, and the initial power is transmitted by the BS. After multiple reflections between the MT and BS, the system eventually reaches resonance, and the signal is fed into the BS for DOA estimation. In contrast, BFLS transmits the signal directly to the BS through beamforming from the MT. When the positioning distance gradually increases, the transmission efficiency of MRLS decreases, and the signal power reflected back by the MT also gradually decreases. At a distance of 6 meters, even when resonance is reached, the signal power ultimately reflected back by the MT is still lower than the actively transmitted signal from the MT to the BS. The RMSE of DOA estimation is inversely related to the SNR, and eventually becomes higher than that of BFLS at 6 m.
\begin{figure*}[!t]
\centering
\subfigure[]{
	\includegraphics[width=0.35\linewidth]{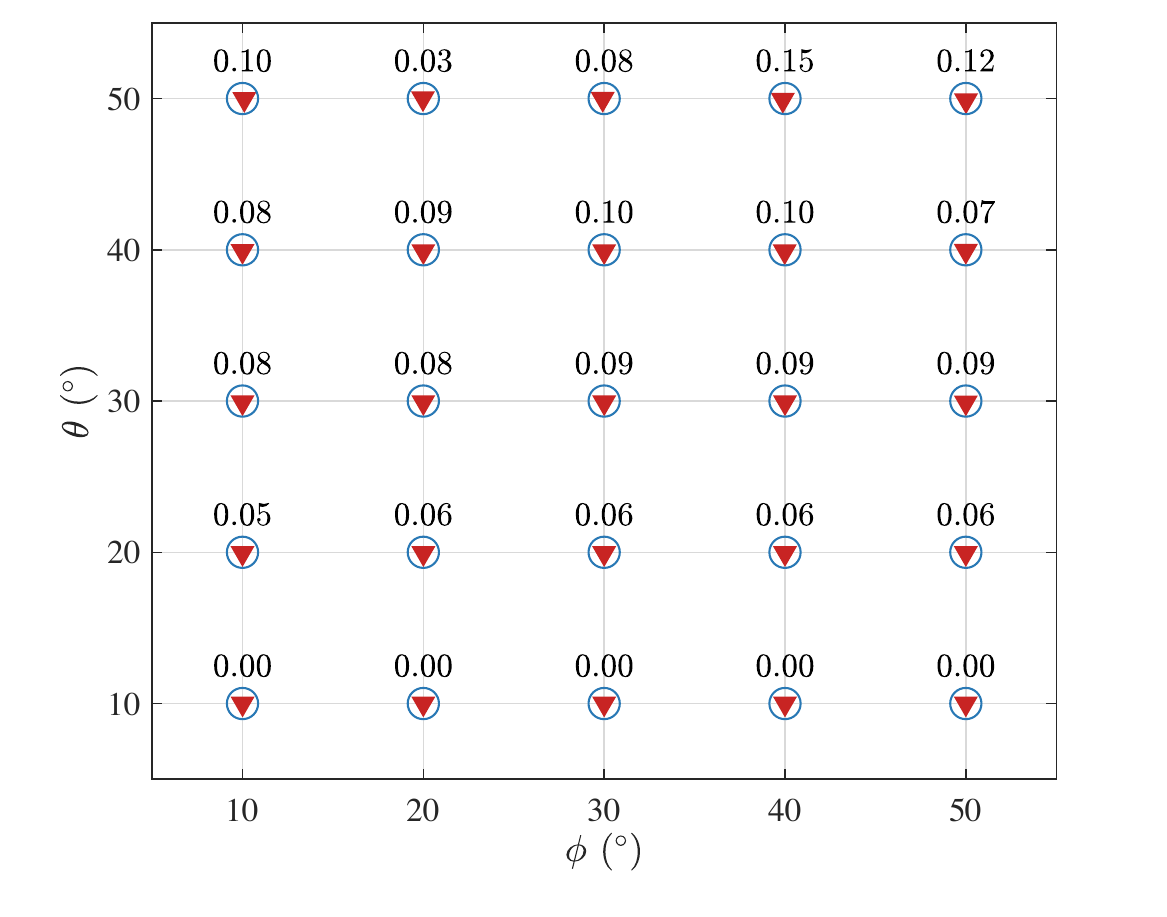}}
\subfigure[]{
	\includegraphics[width=0.35\linewidth]{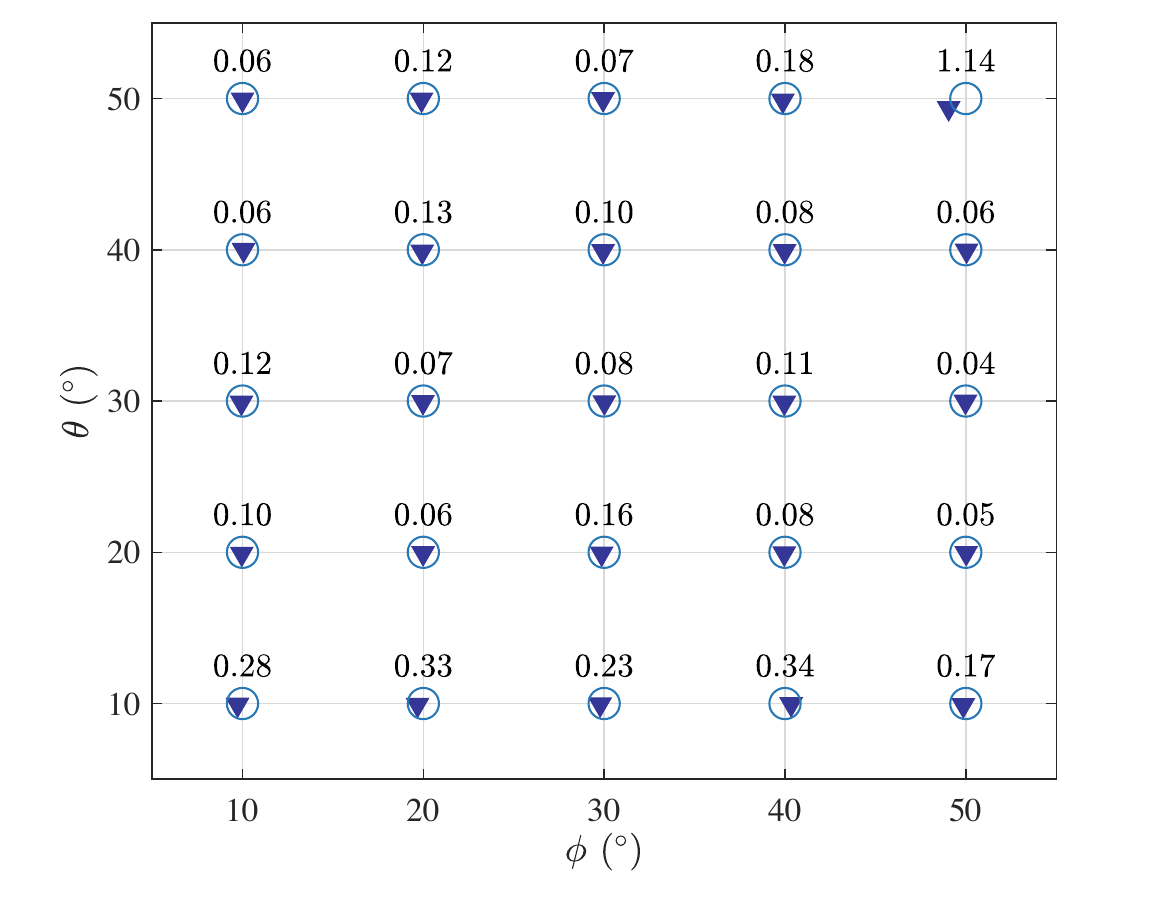}}
 \subfigure[]{
	\includegraphics[width=0.35\linewidth]{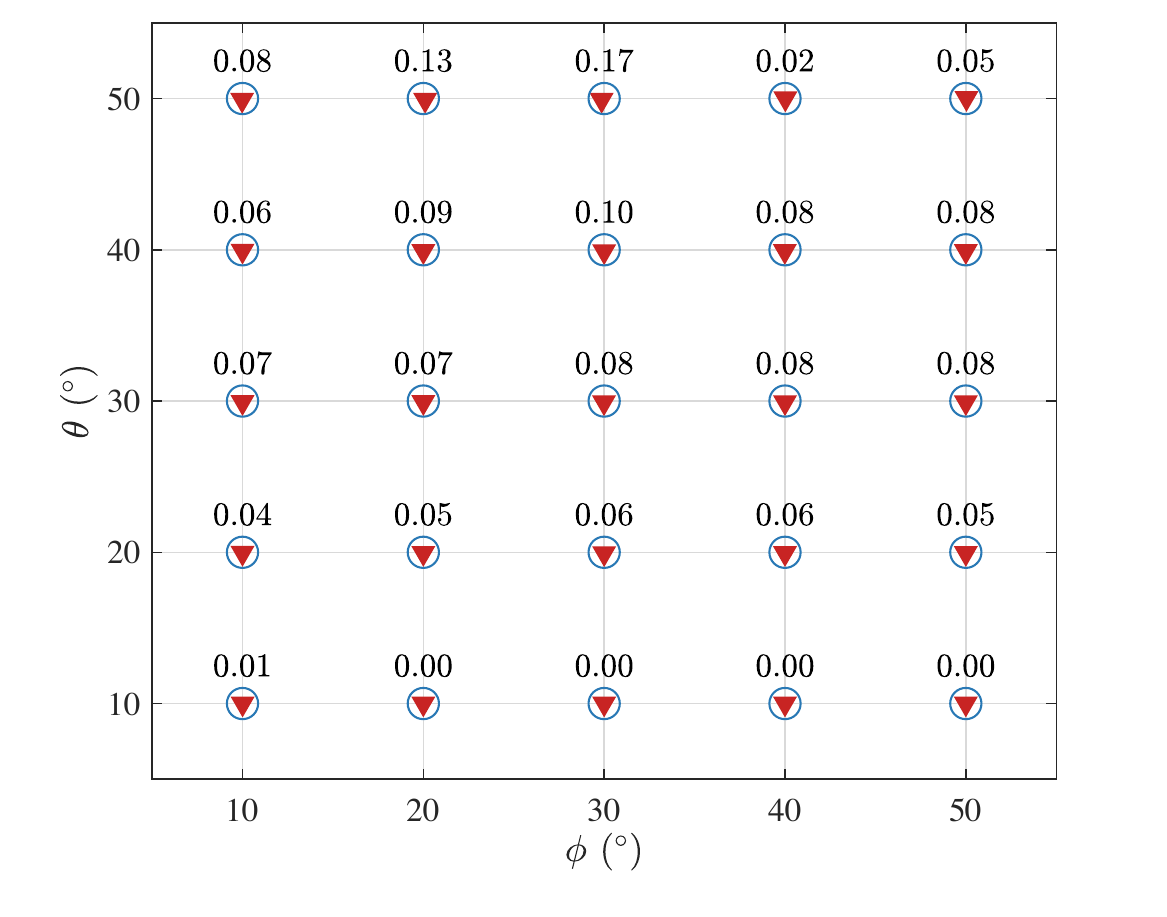}}
        \subfigure[]{
	\includegraphics[width=0.35\linewidth]{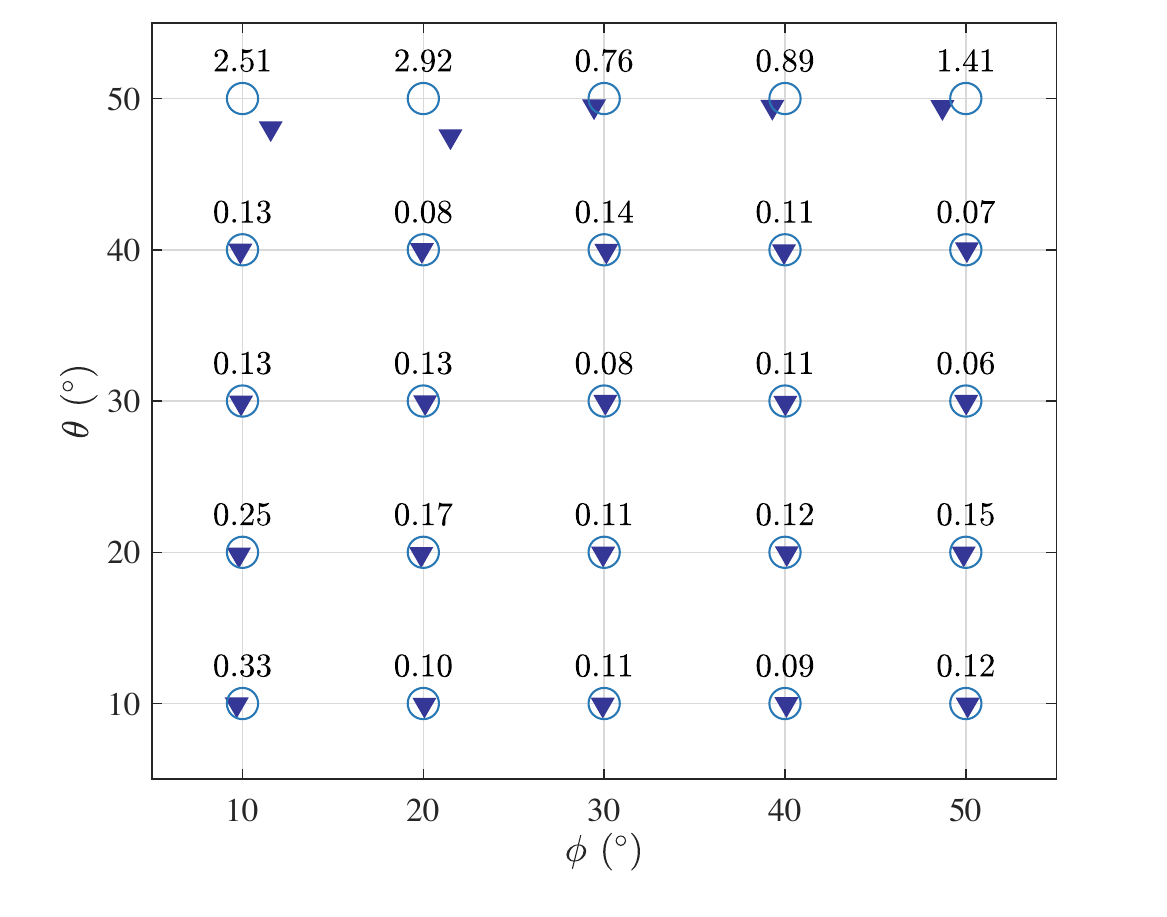}}
 
\caption{The DOA estimation results from different angles. (a) The MRLS estimation results under 0.02 mW noise power; (b) The BFLS estimation results under 0.02 mW noise power; (c) The MRLS estimation results under 0.03 mW noise power; (d) The BFLS estimation results under 0.03 mW noise power.}
\label{DOAVSerror}
\end{figure*}

Figure~\ref{DOAVSerror} shows the DOA estimation results of MRLS and BFLS for 5$\times$5 reference points with distances of 3 m, and elevation and azimuth angles ranging from 10$^\circ$ to 50$^\circ$, under identical noise power conditions. The DOA estimation results of MRLS are shown in red, while those of BFLS are in blue. We conducted 50 Monte Carlo simulations, and the RMSE is annotated in the figure. Figs.~\ref{DOAVSerror}(a) and (b) show the DOA estimation results under a noise power of 0.02 mW, where the DOA errors of both systems are not significantly different. Therefore, we increased the noise power to 0.03 mW, resulting in Figs.~\ref{DOAVSerror}(c) and (d). At this point, it can be clearly seen that the DOA estimation error of MRLS remains very small, whereas BFLS shows significant errors at high elevation angles. This comparison clearly shows that MRLS provides more accurate DOA estimation for MTs in this range of angles compared to BFLS.

\section{Conclusion}
This paper proposes a novel multi-target passive localization scheme. Firstly, a multi-target resonance system structure suitable for the radio frequency band was proposed, and a power cycling model which is between a single BS and multiple MTs, was established within this structure. Then, according to the resonance principle, the BS and multiple MTs simultaneously establish bidirectional echo resonance without the MTs needing to emit additional signals.
Finally, after resonance is formed, the BS estimates the DOA of MTs based on the fusion signals received from them. Numerical evaluations demonstrate that our system can achieve highly directional beam alignment with multiple MTs simultaneously without beam control and can provide higher accuracy in DOA estimation compared to active BFLS. This is achieved without the need for active MTs signal transmission and works effectively within an elevation angle of -50 to 50 degrees at a distance of 6 m, showing promise for applications requiring precise localization.
\bibliographystyle{IEEEtran}

\bibliography{Mybib}

\end{document}